# YOUNG Star detrending for Transiting Exoplanet Recovery (YOUNGSTER) II: Using Self-Organising Maps to explore young star variability in Sectors 1-13 of *TESS* data

Matthew P. Battley,[1,2]* David J. Armstrong[1,2] Don Pollacco,[1,2]

[1]*Dept. of Physics, University of Warwick, Gibbet Hill Road, Coventry CV4 7AL, UK*
[2]*Centre for Exoplanets and Habitability, University of Warwick, Gibbet Hill Road, Coventry CV4 7AL, UK*

## ABSTRACT

Young exoplanets and their corresponding host stars are fascinating laboratories for constraining the timescale of planetary evolution and planet-star interactions. However, because young stars are typically much more active than the older population, in order to discover more young exoplanets, greater knowledge of the wide array of young star variability is needed. Here Kohonen Self Organising Maps (SOMs) are used to explore young star variability present in the first year of observations from the *Transiting Exoplanet Survey Satellite (TESS)*, with such knowledge valuable to perform targeted detrending of young stars in the future. This technique was found to be particularly effective at separating the signals of young eclipsing binaries and potential transiting objects from stellar variability, a list of which are provided in this paper. The effect of pre-training the Self-Organising Maps on known variability classes was tested, but found to be challenging without a significant training set from *TESS*. SOMs were also found to provide an intuitive and informative overview of leftover systematics in the *TESS* data, providing an important new way to characterise troublesome systematics in photometric data-sets. This paper represents the first stage of the wider YOUNGSTER program, which will use a machine-learning-based approach to classification and targeted detrending of young stars in order to improve the recovery of smaller young exoplanets.

## 1 INTRODUCTION

### 1.1 The Young Star Challenge

Young stars and their corresponding exoplanetary systems are equal parts fascinating and troublesome. Existing at epochs where system evolution is ongoing, young exoplanets[1] may experience phenomena such as continuing formation and accretion (Marley et al. 2007; Manara et al. 2019), disk migration (Nelson 2018), dynamical interactions (Ida & Lin 2010; Schlichting et al. 2015) and atmospheric loss (Baraffe et al. 2003; Owen 2019). Unfortunately, these same processes which make young stars such interesting objects to study also vastly increase the difficulty of finding new young exoplanets, especially when coupled with the intrinsically increased stellar activity/variability of young stars (e.g. Skumanich 1972; Soderblom et al. 1991; Mamajek & Hillenbrand 2008; Briceno et al. 2019; Kiman et al. 2021). Particularly challenging (and interesting) are variability types such as evolving spot-induced modulation (Sergi-son 2015), signatures of atmospheric loss and extended atmospheres (Owen 2019), dippers and bursters (Hedges et al. 2018) and young eclipsing binaries (Smith et al. 2021).

However, despite this stellar activity, a few score young transiting exoplanets have now been discovered, with discoveries accelerating in the past decade thanks to the *Kepler* (Borucki et al. 2010), *K2* (Howell et al. 2014) and *Transiting Exoplanet Survey Satellite* (*TESS*, Ricker et al. 2014) missions. A number of different groups have been involved in the discovery of these planets, but particularly prolific have been the ZEIT (*Zodiacal Exoplanets In Time*, e.g. Mann et al. 2016; Rizzuto et al. 2017, 2018) and *THYME* teams (*TESS Hunt for Young and Maturing Exoplanets,* Newton et al. 2019; Rizzuto et al. 2020; Mann et al. 2020; Newton et al. 2021; Tofflemire et al. 2021). The all-sky nature of *TESS*'s mission makes it an attractive instrument for revealing more of these interesting young planets, so a number of further searches are ongoing, often linked to light-curve extraction methods like the Cluster Difference Imaging Photometric Survey (CDIPS, Bouma et al. 2019, 2020), PSF-based Approach to TESS High quality data Of Stellar clusters (PATHOS,e.g. Nardiello et al. 2019, 2020, 2021) and MIT Quick-look Pipeline (QLP, Huang et al. 2020).

---

* E-mail: Matthew.Battley@warwick.ac.uk
[1] n.b. for the purpose of this paper, 'young' is defined as <1Gyr in age.

One of the main challenges hindering further young planet discoveries is the wide variety in photometric variability of young stars, as was explored in Battley et al. (2020). This provides a significant challenge to traditional 'one-size-fits-most' detrending approaches, where often a single window size is chosen for simplicity. Indeed, as Hippke et al. (2019) illustrated in their Wotan detrending paper, most general detrending algorithms recovered only 37% of the $0.5R_{Jup}$ sized young planetary signals injected into young star light-curves, compared to approximately 99% of those injected into an older population. Battley et al. (2020) and Nardiello et al. (2021) recently broke this down further, illustrating how the recovery of planets injected into young-star light-curves varied with planetary radius/radius ratio and period, as well as stellar magnitude and rotation period. In both these cases recovery was seen to drop considerably when injected planetary radii fell below the Hot Jupiter size regime.

In order to push down to smaller and younger exoplanets, it is crucial to both gain a deeper understanding of the variability present in these young stars and to work out the most effective ways of removing it. With increased knowledge of each potential host star's variability, each target could then be detrended using a method best suited to its variability type. This maximises the ability to remove complex variability without also removing transits.

Large-scale stellar variability classification has been a subject of interest since the first large-scale surveys such as the Hipparcos mission[2], Optical Gravitational Lensing Experiment (OGLE[3]), and All-Sky Automated Survey (ASAS[4]) began making traditional human eyeballing untenable. Machine-learning methods are typically grouped into 'supervised' and 'unsupervised' methods, depending whether they are provided with labelled training data or not. Most of the earliest variability classifiers focused on supervised methods, labelling distinct variability groups and 'features' before applying versions of neural networks (Pojmanski 2002) or decision trees (Ball et al. 2006). Similarly, Debosscher et al. (2007), Sarro et al. (2009) and Debosscher et al. (2011) built a pipeline based on Gaussian Mixtures and Bayesian networks to conduct global variability classification (and training) on data from CoRoT (Auvergne et al. 2009), OGLE and Kepler Quarter 1. Richards et al. (2011) returned to the decision-tree (or Random Forest) method to classify stars from OGLE and Hipparcos, eventually using these as a training set to construct the large Machine-learned ASAS Classification Catalog (MACC Richards et al. 2012). The versatility of retraining supervised nachine learning methods have also allowed them to be applied to slightly more unusual types of variability, including for the discovery of transients in synoptic survey imaging data (Brink et al. 2013), detecting anomalous light curves in massive astronomical catalogs (Nun et al. 2014) and even on MACHO (MAssive Compact Halo Objects) and LINEAR (LincoIn Near-Earth Asteroid Research) data Kim & Bailer-Jones (2016).

Unsupervised methods on the other hand allow their algorithms to discover information and clusters in the data without pre-decided labels or groups. This can lead to new data groups emerging and is useful for situations in which not all potential solutions are known. However, these methods are comparatively rare in stellar classification analyses. One of the earliest applications of such a method was completed by (Eyer & Blake 2005), using an unsupervised Bayesian classifier on features from Fourier-space to classify variable stars in ASAS data releases 1-2. In another direction, Valenzuela & Pichara (2018) created a 'Variability Tree' which allowed simultaneous feature generation and varaibility search to be completed concurrently on OGLE, MACHO and Kepler data. Meanwhile others such as Modak et al. (2020) and (Armstrong et al. 2016) have turned to clustering methods instead, allowing similar variability shapes or light-curves to be places in similar clusters.

More recently, there has been a drive to combine the different benefits provided by both unsupervised and supervised methods into a single classification. Naul et al. (2018) for example combined an unsupervised autoencoded recurrent neural network to find features within their dataset with a supervised random forest to make the final classifications for optical variable star catalogs. Meanwhile the large *TESS* Data for Asteroseismology (T'DA) collaboration is currently working on a large stellar variability classification pipeline (Audenaert et al. 2021), or 'meta-classifier', which combines the Multiclass Solar-like Oscillation Shape Hunter (multiSLOSH Hon et al. 2018) neural network, Random Forest General Classification (RFGC, a hybrid unsupervised Self-Organising Map and Random Forest implementation developed by Armstrong et al. 2016), Supervised Random Forest Variability Classifier Using High-resolution Photometry Attributes in TESS Data (SORTING-HAT Audenaert et al. 2021) and Gradient Boosting General Classification (GBGC Friedman 2001) methods. Furthermore, the *Gaia* mission collaboration applies a wide variety of classification pipelines depending on the type of variability, with classifications continuing to be improved with the ongoing mission (Gaia Collaboration et al. 2016, 2018b; Eyer et al. 2019)

However, despite these widespread approaches to general stellar variability analyses, machine-learning classification of variability in young stars remains in comparative infancy. Due to challenges like swiftly evolving and unusually shaped activity, coupled with a traditional dearth of bona-fide young stars (Gagné et al. 2018a), most classification of young star light-curves has been based on human eyeballing (e.g. Cody & Hillenbrand 2018). However, as young stars exhibit a breadth and depth of stellar variability often not seen in older populations, there is significant value in completing variability classification independent of the older population. Thankfully the improved performance and coverage of the new allsky *TESS* and *Gaia* missions is now beginning to supply the data necessary for young-star specific analyses, making this a useful time to begin such analyses. Indeed Hedges et al. (2018) have recently shown the power of young-star specific classification in a search for for dippers/bursters found from supervised machine learning.

### 1.2 The YOUNGSTER Program

The aim of the "YOUNG Star detrending for Transiting Exoplanet Recovery" (YOUNGSTER) program, is to develop a targeted detrending approach in young exoplanet searches by using information about the stellar variability of potential young host stars to inform detrending of different categories of variability. Unlike a traditional 'one size-fits-most' detrending approach, this program relies on first classifying different types of young star variability which can then be treated independently. By considering the variability of potential host stars during the analysis (rather than simply eliminating it), this approach helps to reduce the prevalence of leftover signals induced by improper detrending and should aid the search for shallower exoplanets around young stars considerably. Furthermore, it provides additional information about the individual systems, thus aiding further analysis of the system as a whole.

The first paper in this series (Battley et al. 2020, hereafter YOUNGSTER I) presented a new lowess-based detrending method which showed promise as a dedicated young-star detrending method, but also highlighted some of the challenging variability classes such as very short-period rotation (<1.5 day rotation periods). This paper, paper II, presents the results from the use of an unsupervised Self-Organising Map (SOM, Kohonen 1982, 2001) as a tool to explore young star variability further, and acts as a first step into more targeted characterisation and sorting of young star variability for exoplanet searches. It also provides a useful additional method of recognising and removing leftover systematics in young star light-curves and gives an efficient method to recover example light-curves of unusual variability types on which to train future classification methods. Paper III (Battley et al., in prep) will extend this classification using a supervised random-forest implementation which combines additional light-curve statistics with the results from the SOM in order to give final classifications for each star. Specifically, it will combine the SOM position and Euclidean distance of each light-curve from its closest SOM pixel with noise statistics and (a)periodic information about the light-curves in order to give a final classification for each light-curve. Future papers in this series will then use these classifications as test-beds to develop new targeted detrending methods and apply these to each variability group in order to search for smaller transiting exoplanets.

### 1.3 Self-Organising Maps

Self-Organising Maps (SOMs; aka Self-Organizing Maps) are an unsupervised machine-learning technique originally developed by Kohonen (1982, 2001) which allows for sorting of data based on topology, or similar shapes. A form of artificial neural-network birthed from research into neural networks in the human brain (Kohonen 1982), SOMs use short-range lateral feedback to group similar input data into similar regions of a 'map'. The final map, usually 1-2 dimensions for clarity, can then be used to explore groups of similar data or to more simply investigate the evolution of a complex data-set in a form of dimensionality reduction (Armstrong et al. 2016). One key advantage of Self-Organising Maps is their unsupervised nature, meaning that underlying trends and groups in the input data can be explored without the need for pre-allocated classes of variability or significant knowledge about the target stars. This makes them particularly useful for young star studies, where new classes of photometric variability are still being discovered and understood (e.g. Zieba et al. 2019; Tajiri et al. 2020). A useful extension to this behaviour is that not only are those data which are most similar placed closest together in the final map, but those which are most different are placed furthest apart (Mostert et al. 2021). However, an important proviso with the use of the SOM is that far more importance should be given to the relative position of the analysed data compared to its overall distribution, since, as noted by Geach (2012), the randomness of the initial SOM setup results in a different final distribution on each new run, even for precisely the same input data-set. This means that while the same input data should always end up closest to those data which are most similar, the overall shape of the SOM may be different in each run depending on how the som was initialised in each instance.

SOMs are particularly popular in computer science, quantitative biology and statistics (e.g. Rahmani et al. 2019; Burnap et al. 2018), however until recently have been relatively under-utilised in astronomy. Brett et al. (2004) were the first to use this technique for astrophysical data, using self-organising maps to classify a set of synthetic light-curves and 1026 real light-curves from the ROTSE experiment (Akerlof et al. 2000). As well as providing an overview of how the SOM evolves over different numbers of iterations, Brett et al. (2004) illustrated that the performance of the SOM was relatively insensitive to changes in the value and decay shape of the two main SOM parameters: the learning rate $\alpha$ and effective neighbourhood size $\sigma$. This makes it a versatile and easy approach to use for a wide range of data-sets.

More recently Armstrong et al. (2016) extended this work by using a SOM alongside a supervised Random Forest implementation to classify the variability (or lack thereof) of all stars in Campaigns 0-4 of the K2 mission. In this work Armstrong et al. (2016) used a 2D 40x40 Kohonen layer with an initial learning rate of $\alpha_0 = 0.1$ $\sigma_0 = 40$ (the width of the Kohonen layer) to clearly demonstrate the utility of the SOM for splitting types of photometric variability by training the SOM on a set of 4047 known variables from campaigns 0-2 of the *K2* mission. An improved version of this same SOM setup has recently been folded into the wider *TESS* T'DA stellar variability classifier, combined with a small selection of other supervised and unsupervised classification methods (Audenaert et al. 2021), however it has never previously been tested on the unique groups of variability seen in young stars. In related works (Armstrong et al. 2016, 2020) have also used SOMs to classify transit shapes in an effort to aid automated vetting of planetary transits.

In wider astronomy SOMs have been used for a wide array of analyses, from Photometric Redshift calibration (Geach 2012; Carrasco Kind & Brunner 2014; Wright et al. 2020a,b), Radio astronomy (Galvin et al. 2020; Mostert et al. 2021) and spectroscopic image segmentation (Fustes et al. 2013; Schilliro & Romano 2021) to classification of Galaxies (Naim et al. 1997; Johnston et al. 2021), AGNs (Faisst et al. 2019) and unusual quasars (Meusinger et al. 2012). Of particular note were the discoveries of Johnston et al. (2021) who showed how effective SOMs can be at identifying systematics alongside the main data-groups (in their case in a study of galaxy clustering in the Kilo-Degree Survey), and Khacef et al. (2020)'s suggestion that the efficiency of SOMs could be improved for very large data-sets by applying the SOM to extracted features rather than the raw data. Although many of these wider applications are less relevant to the current work, the wide range of uses further highlights the versatility of the SOM for complex data, alongside the ability to classify a wide array of interesting behaviour into groups. This versatility, coupled with the unsupervised clustering algorithm, intuitive results format and promising track record for photometric variability classification make the Self-Organising Map an appealing technique for exploring and classifying young star variability.

## 2 LIGHT-CURVE SOURCE AND TARGET SELECTION

### 2.1 Light-curve extraction

For machine-learning approaches like self-organising maps, two factors are crucial: consistency and data volume. Now, in the era of the *Transiting Exoplanet Survey Satellite* (*TESS*, Ricker et al. 2014), the community has access to a uniform source of millions of light-curves from an ever-increasing proportion of the sky, including hundreds of thousands of young star light-curves extracted from the Full-Frame-Image (FFI) data. Numerous different sources of FFI light-curves are now available, including a number of very large data-sources like the *TESS* SPOC (Caldwell et al. 2020a), MIT Quick-Look (QLP, Huang et al. 2020) and *eleanor* pipelines (Feinstein et al. 2019). Crucially for this analysis, there are now also a selection of dedicated young star light-curve extraction programs such as the CDIPS (Bouma et al. 2019), PATHOS (Nardiello 2020), and Oelkers & Stassun (2018)'s Difference Imaging Analysis light-curves. These specialised extraction pipelines are particularly useful for dense regions where young stars are commonly found due to their difference imaging (n.b. also used to extract Huang et al. (2020)'s QLP light-curves) or PSF-based approaches.

In this work, light-curves extracted from the Cluster Difference Imaging Photometric Survey (CDIPS) pipeline (Bouma et al. 2019; Bhatti et al. 2019)[5] were chosen as the primary data source due to their dedicated young-star extraction approach and the availability of a large number of light-curves for the entire first year of *TESS*'s Primary Mission (Sectors 1-13). They also provide a 'lightly detrended' PCA form of light-curve particularly useful for the SOM technique, as discussed later. Through comparison to different year-1 light-curve sources currently available to the community, this form of light-curve was found to best preserve signals of stellar activity whilst removing additional spacecraft systematics. Note however that because no light-curves from the second year of *TESS*'s observations were available from the CDIPS pipeline at the time of writing, the more widely-available QLP light-curves of (Huang et al. 2020) were chosen as a backup data source for exploring scattered light systematics in section 4.2.

The CDIPS pipeline uses a difference-imaging approach to extract light-curves from the *TESS* FFIs. Based largely on the FITSH package of Pál (2012), the pipeline is similar in vein to other difference imaging pipelines designed for use with the *TESS* data, such as that of Oelkers & Stassun (2018) and Huang et al. (2020). An in-depth description of the full CDIPS pipeline can be found in Bouma et al. (2019), but is briefly explained here. The pipeline begins with the calibrated *TESS* SPOC FFIs, performs large-scale background subtraction and then builds an astrometric reference frame based on a series of the brightest stars in each image, verified with the World Coordinate system (WCS) and the Gaia-DR2 catalog (Gaia Collaboration et al. 2016, 2018b). All calibrated images are then transformed to this frame before a photometric reference frame is constructed from 50 images with low background noise. In order to get the final light-curves, each target frame is then subtracted from the photometric reference frame and aperture photometry applied to each source.

It should be noted that while times of known momentum dumps and coarse pointing were removed using the *TESS* data release notes[6] early in the CDIPS pipeline, a number of additional systematics from the spacecraft motion and scattered light still exist in the data after the main difference-imaging extraction. In order to combat this, Bouma et al. (2019) present two partial-detrending approaches: Principal Component Analysis (PCA) and the Trend-Filtering Algorithm (TFA). Both approaches rely on ensemble-detrending of the light-curve of interest using up to 200 nearby 'template' stars without significant periodicity. Principal Component Analysis works by deriving 10-15 principal components which comprise the systematic trends in each individual CCD and then building a model for each light-curve using a subset of these components, weighted by consideration of linear least squares. This removes the largest systematics on each CCD while avoiding over-fitting of most stellar variabil-ity. The specific implementation used here was that included in the SCIKIT-LEARN package (Pedregosa et al. 2011). Meanwhile TFA works by deriving a filter function to remove long-period trends and thus is well-suited to transit studies, but too severe for analysis of stellar signals, as it often removes/distorts interesting stellar variability alongside true systematics.

Due to its ability to remove spacecraft systematics whilst maintaining most stellar variability, the PCA detrended light-curves were eventually chosen as the primary brightness data-source for this analysis. For consistency, aperture 2 (1.5 pixels in radius) was used for all light-curves, chosen as a balance of noise and the relative crowding of many young stars. The final data vectors used for each object were thus time = '*TMID_BJD*', magnitude = '*PCA2*' and mag_err = '*IRE2*'. Finally, these magnitudes were converted to normalised flux to prepare them for further analysis, based on the median flux for each source.

## 2.2  Target Selection

An additional benefit of using the CDIPS light-curves is that all light-curves extracted by the pipeline show evidence of youth, either through candidate membership of young clusters/associations/moving groups or due to external photometric indications. The CDIPS target list was constructed from four large archival catalogs of open clusters (Cantat-Gaudin et al. 2018; Gaia Collaboration et al. 2018b,a; Kharchenko et al. 2013; Dias et al. 2014) as well as nine smaller catalogs of moving groups and stellar associations (Gagné et al. 2018a,b; Gagné & Faherty 2018; Kraus et al. 2014; Röser et al. 2011; Bell et al. 2017; Rizzuto et al. 2011; Oh et al. 2017; Zari et al. 2019). This resulted in an overall target list of 1,061,447 stars, for which 671,894 light-curves are available over the first year of *TESS* observations.[7] All of these light-curves were included in the present analysis. For a full description of the target list and its construction see Bouma et al. (2019). However, some additional care is needed when using this list, as Bouma et al. (2019) admit that the it was compiled for "completeness, not accuracy", so the true youth of some of the included objects should be treated with caution. Nonetheless, since the overall aim of the YOUNGSTER program is to understand and detrend different types of variability in the wider sample of young stars, the inclusion of some older, likely less-active sources is of little concern.

## 3  PERIOD-FINDING AND SOM ARRAY PREPARATION

The first step in characterising the variability of the young star sample is deriving the variability period of each source. Although most instrumental systematic effects were removed in the CDIPS and PCA steps, in order to combat the common systematic period of 13.5 days (half the length of a typical *TESS* sector), a 13.5 day period ($f = 1/14$) sinusoid of the form

$$y = a \sin 2\pi f t + b \cos 2\pi f t + c \qquad (1)$$

was fitted to the data and removed by division from the flux data before the main period search. Primary, secondary and tertiary periods for each light-curve were then found using the common Lomb-Scargle method (Lomb 1976; Scargle 1982), as implemented into python as the LOMBSCARGLE function in the python ASTROPY library[8] (The Astropy Collaboration et al. 2013; Astropy Collaboration et al. 2018). These periods were found iteratively, with the primary period in each iteration obtained using the Lomb-Scargle method and then removed by fitting a sinusoid of the same form used for the systematic 13.5d period. As the present work is most interested in the overall shape of the variability rather than the absolute period for each target, the periods were allowed to vary from 0.25 days (4 $d^{-1}$) to 25 days (0.04 $d^{-1}$), allowing the non-variable or aperiodic light-curves to remain largely unfolded whilst not risking the introduction of more noise-related frequencies at much longer frequencies. Given that most periods found were on the order of multiple days, extending the upper frequency to the Nyquist frequency (∼1h period) is not expected to change this analysis significantly. Furthermore, attempting to remove stellar activity with periods shorter than a quarter of a day would be exceedingly difficult in the 30min data.

To prepare for the SOM analysis, the extracted CDIPS PCA light-curves were folded by their strongest period and normalised between 0 and 1 to allow direct comparisons of different shapes of variability. Similar to Armstrong et al. (2016), these phase-curves were binned into 64 bins in order to save computational time whilst maintaining the overall shape of the variability. The phased arrays were then aligned such that the minimum point always occurred as the first element in the array, before all phase-curves were arranged into per-sector lists as the main input to the SOM analysis. Where gaps existed in the light-curves of over 0.5 days, these were filled in with linear interpolation (after the Lomb-Scargle period search) in order to avoid processing issues. In most sectors this was only required across the 1.1 day inter-orbital gap, and thus was not found to significantly affect the shape of the binned and folded phase curve. As an additional test, phase arrays for all light-curves from Sector 2 (except those with periods resulting in gaps in the phase-curve) were reconstructed with the gap retained instead of interpolation and mapped onto the final SOMs trained on Sectors 1-5 (Figure 2 and 1-13 (Figure 11 respectively. All light-curves in well-defined variability clusters (such as EA/EB eclipsing binaries) were found to fall precisely on their original positions in both test cases, with 82-86% of all light-curves falling within 10 pixels of their original position (and ~70% overall within 2 pixels). As the shape associated with each Kohonen pixel in the more amorphous region of the SOM was not observed to evolve quickly over the course of 10 pixels, the imposed linear interpolation does not appear to result in significant mis-classifications of shape.

## 4 SELF ORGANISING MAP EVOLUTION

### 4.1 SOM Setup and Algorithm

The SOM implementation used in this work is similar to the one used by Brett et al. (2004) and Armstrong et al. (2016), but is briefly explained here for context. The SOM works by minimising the Euclidean distance between each phase curve and the 'Best-Matching-Unit' (BMU) within the Kohonen Layer. Three main parameters are important for the setup of the SOM: the initial learning rate, $\alpha_0$, initial learning radius, $\sigma_0$ and the SOM shape. The learning rate controls how quickly pixels in the Kohonen layer are changed, while the learning radius controls how large the neighbourhood of comparison is during each iteration. Brett et al. (2004) recommended learning rate values of $\alpha_0 \leq 0.2$ in order to prevent undesirable movement and constant restructuring with further iterations, however the specific value of the learning radius was not found to have a significant effect on SOM performance. Similar to Armstrong et al. (2016), these parameters were eventually chosen as $\alpha_0 = 0.1$ and $\sigma_0 = r$, where $r$ is the SOM radius. A 'radius' is defined here instead of a length, because the boundaries of the SOM are cyclic, wrapping around at each edge. Also following Armstrong et al. (2016), the nominal SOM shape was set to a 2D square of 40x40 pixels for a good balance of detail and required human eyeballing time. However, variations on this shape are also explored in Section 5.2.

Once the input parameters have been chosen, the algorithm proceeds as follows. Firstly, each pixel in the Kohonen layer is randomised such that the 64 elements within each one all have values between 0 and 1. For each iteration, each phase curve array (prepared in Section 3) is compared to the Kohonen layer to find its BMU, by minimising the Euclidean distance between the pixel elements and the phased array elements. Every element in each pixel of the Kohonen layer is then updated according to the expression

$$m_{xy,k,new} = m_{xy,k,old} + \alpha e^{-d_{xy}^2/2\sigma^2}(s_k - m_{xy,k,old}) \quad (2)$$

following the formalism of (Armstrong et al. 2016), where $m_{xy,k}$ is the value $m$ of the pixel at coordinates $x, y$, element $k$ in the phase curve, $d_{xy}$ is the Euclidean distance of that pixel from the BMU in the layer, and $s_k$ is the $k$th element of the input phase curve currently being compared. This process is repeated for each iteration, with learning rate and learning radius evolving as follows

$$\alpha = \alpha_0(1 - \frac{i}{n_{iter}}) \quad (3)$$

$$\sigma = \sigma_0 e^{(\frac{-i log(r)}{n_{iter}})} \quad (4)$$

where $i$ is the iteration number, $r$ is the radius of the Kohonen layer and $n_{iter}$ is the total number of iterations (chosen as 1000 in this work). Note that the SOM performance is relatively insensitive to the precise functional form of how $\alpha$ and $\sigma$ evolve, as explored by Brett et al. (2004).

The final result of the algorithm is the Kohonen map, a representation of the overall variability present in the data-set. In this case each pixel effectively represents an average of those phase arrays which ended up at it after the final iteration. Querying individual pixels in the Kohonen map thus gives an approximation of each phase-curve at that pixel.

The specific software implementation used in this work was adapted from the TRANSITSOM code of (Armstrong et al. 2017)[9], which in turn was based on the SOM code from the PYMVPA PYTHON package (Hanke et al., 2009).[10]

### 4.2 Identification and removal of systematics

One of the additional benefits of the unsupervised clustering method used by the SOM is the easy identification of leftover systematics in the *TESS* light-curves. Indeed Geach (2012) refer to the SOM as a form of "non-linear principal component analysis" due to its ability to find patterns and common features in data. The intuitive and visual results interface offered by the SOM makes understanding these systematics simple and efficient. To explore this aspect of SOM

---

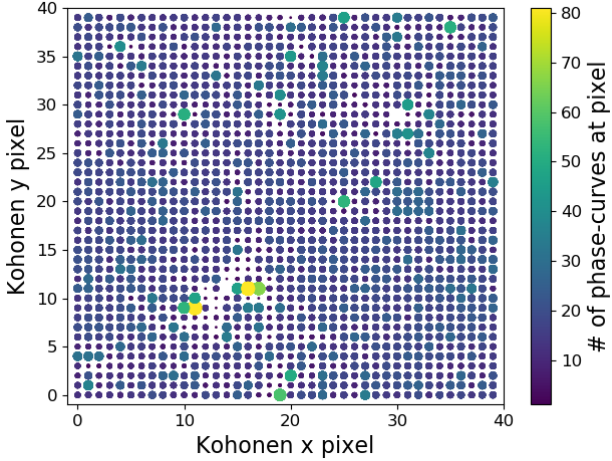

**Figure 1.** Original S1-5 SOM, constructed from CDIPS 'IFL2' flux light-curves. The Kohonen x and y pixels correspond to x and y in Equation 2, with individual pixels referred to in the form [x, y]. Both colour (low = purple; high = yellow) and pixel radius are a function of the number of phase-curves situated at that pixel in the final SOM map. The larger, brighter pixels thus represent the pixels which the greatest number of phase-arrays ended up at, corresponding to the most common shapes in the input light-curve data.

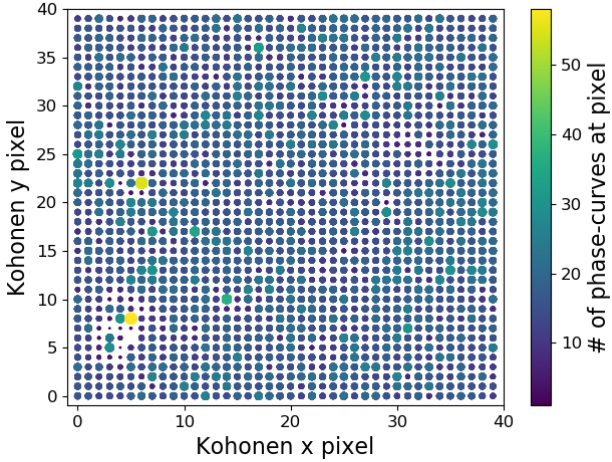

**Figure 2.** Final S1-5 SOM using PCA flux. Colours and pixel size share same meaning as in Figure 1. In this case a more even distribution is present than in Figure 1, but clear hot-spots can be seen at pixels [x,y] = [3,5] and [5,8], representing EA and EB-like eclipsing binary signals.

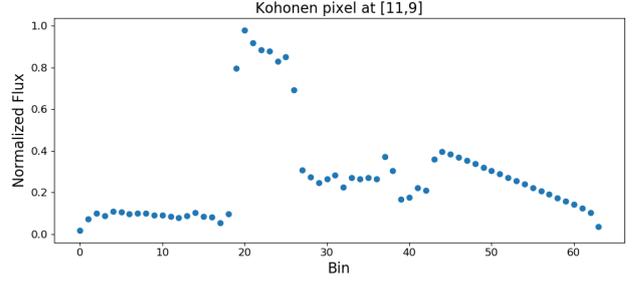

**Figure 3.** The most common Kohonen pixel [11,9] in the CDIPS S1-5 'IFL2' flux run, showing clear flux-jump around bin 20 in the phase-curve. Note that 'Bin' here is synonymous with phase, with bin 0 at phase = 0 and bin 63 at phase = 1.

.

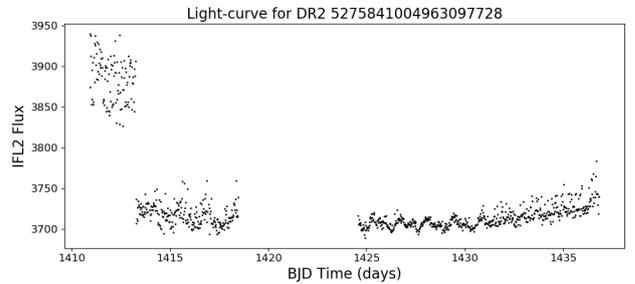

**Figure 4.** IFL2 light-curve for young star Gaia DR2 5275841004963097728, representative of all those light-curves with SOM arrays ending up at pixel [11,9] in the IFL2 SOM for CDIPS S1-5. The clear systematic seen in Figure 3 is explained by a flux jump present in the first part of the first orbit, a common systematic in Sector 4.

performance, SOMs generated for the raw flux ('IFL2') and PCA ('PCA2') CDIPS light-curves for Sectors 1-5 (shown in Figures 1 and 2 respectively) are compared here.

While the SOM based on PCA flux (Figure 2) exhibited a relatively even distribution of phase-curves aside from some 'hot-spots' of large numbers of eclipsing-binary-like light-curves near pixels [x,y] = [3,5] and [5,8] (as is common for all sectors in the final SOM - see main discussion in Section 5), the SOM based on 'IFL2' fluxes (Figure 1) extracted straight from the difference images exhibited considerably more hot-spots, spread over much of the 40x40 grid. The most common pixels for phase curves to end up at in the 'IFL2' SOM were pixels [11,9] and [16,11]. When these pixels were queried in the Kohonen layer, all of the phase curves clustered at these points were found to be from Sector 4, with a very distinctive SOM array shown in Figure 3. At first glance this phase-array appears to represent interesting outburst-like activity, however investigating this feature further by checking the raw light-curves upon which the phase-curves were based (e.g. for object Gaia DR2 5275841004963097728 shown in Figure 4) revealed that the cause of this unusual shape was simply a common flux-jump present in the first part of many dimmer Sector 4 light-curves. Variations in the phase of this flux-jump in the final phase-curves accounted for a large number of the other 'hot-pixels', especially in the regions immediately adjacent to [11,9] and [16,11]. Closer inspection of the wider IFL2-based SOM revealed that other individual hot-pixels could be explained by additional systematics, such as flux ramps (e.g. [10,29], [20,2] and [25,39]) and sudden flux dips, mimicking Algol-type eclipsing binaries (e.g. [31,30] and [31,27]).

While the significance of the leftover systematics in the raw IFL2 flux light-curves varies from sector to sector,[11] the presence of such effects even in a single sector make it concerning for variability analyses, as telling the difference between true variability and leftover systematics shared with adjacent stars is made considerably more difficult. This suggested that some ensemble post-processing of the light-curves was wise, such as that performed for the 'lightly-detrended' PCA light-curves. Indeed, when the PCA light-curves were used instead of the IFL2 fluxes, the final result (Figure 2) was no longer systematics-dominated, with the hottest pixels corresponding to eclipsing binaries [5,8] and 'noisy' sinusoids ([6,22]; sinusoids with significant scatter), alongside the typical evolution of activity discussed in Sector 5. This suggests that these light-curves are far more suited to analyses of overall variability compared to the raw 'IFL2' ones, backing up their choice as the primary light-curve in the full year SOM analysis.

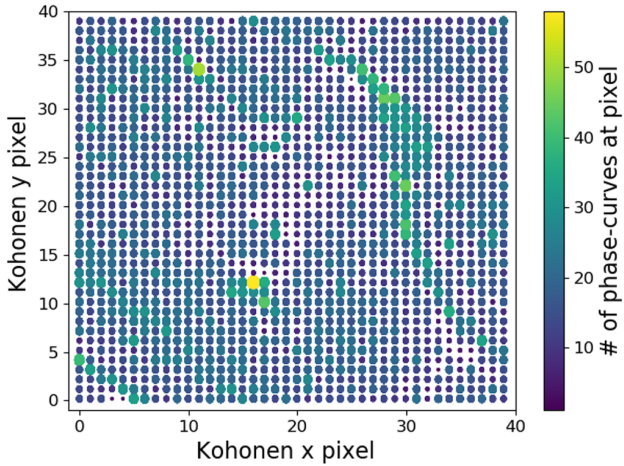

**Figure 5.** Original S17 QLP SOM (after cleaning with the quality flag only) showing scattered light systematics, particularly in the vicinity of [16,12] and [30,27].

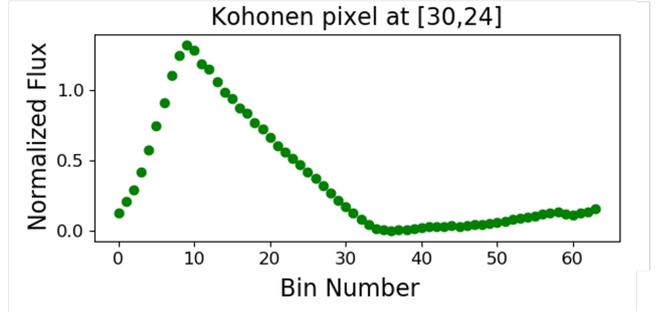

**Figure 6.** Typical scattered light pixel in QLP quality clean light-curves.

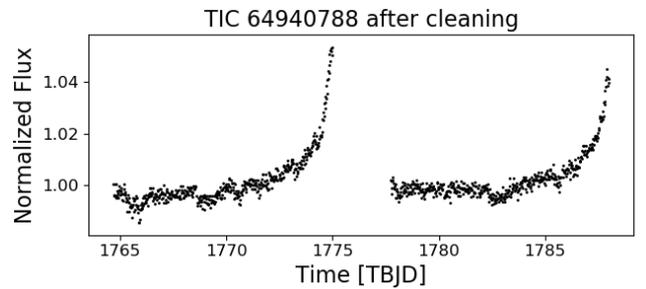

**Figure 7.** Example of scattered light systematics leftover in 'cleaned' QLP light-curve.

Another challenging systematic known in *TESS* data is scattered light, which is particularly prevalent in early Year II of the *TESS* data. Because CDIPS light-curves were only available for Year I at the time of writing, this systematic effect was explored using QLP light-curves instead (Huang et al. 2020). In this case young star targets were chosen from the catalogues upon which the CDIPS target list is based, by performing a cross-match between them and the QLP light-curves. This resulted in 24,495 light-curves at the time of writing. Light-curves were extracted from the QLP light-curves using the 'SAP_FLUX' keyword, before being cleaned by removing any data points with QLP 'QUALITY' values > 4095 from the light-curve. Moments of known poor telescope pointing (as detailed in the *TESS* data release notes[12]) were also removed where these were not caught by the quality cut. This cleaning method resulted in removing similar sections of the data removed in the initial steps of the CDIPS pipeline.

A SOM was then built using all 24,495 light-curves using the same setup outlined in Section 4.1, resulting in the final SOM map shown in Figure 5. What is immediately obvious is the unusually-shaped clusters around [16,12] and [30,27]. These structures display a much wider spatial extent compared to those clusters in the raw IFL2 CDIPS SOM shown in Figure 1, appearing as large collections of light-curves spread over many pixels, instead of single 'hot-pixels' at defined points on the SOM. This implies that they are variations on a shape theme, rather than exactly the same shape each time, hence spread out further on the SOM. Closer inspection of these regions reveal SOM pixels with shapes like that presented in Figure 6, featuring a steep climb and a slow descent in a triangluar pattern. Querying typical light-curves at this pixel makes it clear that this is caused by steep flux-ramps at the end of each orbit. According to the *TESS* data release notes for S17, this was typically caused by scattered light from the Earth and Moon glinting across the detector at these epochs, and can be clearly seen in the background flux (e.g. panel 2 of Figure 9).

Removing the effect of this scattered light is complicated, as it varies in significance depending on a wide range of factors, including position on the ccd/detectors, brightness and crowding on the source star and the angle of the space-craft. These scattered-light systematics were still retained (though slightly reduced in severity) when the most affected detector (Camera 3, Detector 1) was removed from the analysis. Background-based and initial ensemble-based detrending methods based on nearby stars were also trialled, but found to have variable success. The reason for this is best explained by Figure 8, showing the raw light-curves for TIC 64940788 and the 9 closest sources with QLP light-curves from *TESS*. It is immediately clear that removing an average of the activity will not entirely remove the scattered light signal of the original light-curve, and may actually distort it further in some cases. This is a product of the scattered light effect evolving in a complex manner, affected by such factors as the detector position and magnitude of each source. A simple epoch-cut was also found to be challenging, as the epoch where scattered light systematics began to dominate varied between source to source and, in some brighter cases, never to occur at all.

Eventually however, a hybrid approach was found to have some success by considering the background error data row 'SAP_BKG'. Through comparison of the SAP flux ('SAP_FLUX'), SAP background flux ('SAP_BKG') and SAP background error ('SAP_BKG_ERR'), the SAP flux was found to become scattered-light dominated beyond the point in each orbit when the background error exceeded twice the median background err. This relationship is demonstrated visually in Figure 9. Cutting the light-curve at this point was found to remove the majority of the scattered light systematic whilst minimising loss of useful data at the same epochs in objects less affected by scattered light. After cleaning all light-curves based on this background error cut approach, a new SOM was built for the S17 QLP light-curves, which is shown in Figure 10. This SOM no longer displays the clear systematic scattered light effects seen in the original S17 QLP SOM (Figure 5) and instead results in a SOM similar to the final CDIPS SOM for Sectors 1-13 (Figure 11), with steadily evolving stellar variability over the whole 40x40 SOM and hot-spots of eclipsing-binary-like activity around [8,4] and [12,1]. The use of such a cleaning technique thus extends the potential of SOM analysis to the prevalent QLP light-curves as well.

___

[12] https://archive.stsci.edu/tess/tess_drn.html

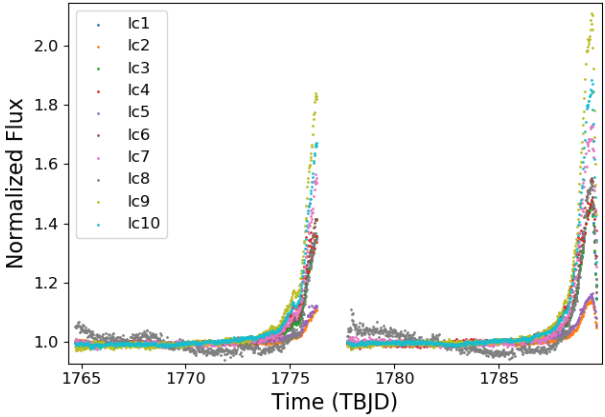

**Figure 8.** Scattered Light variation on 10 adjacent QLP light-curves, illustrating the challenge of removing this systematic effect through simple co-trending techniques

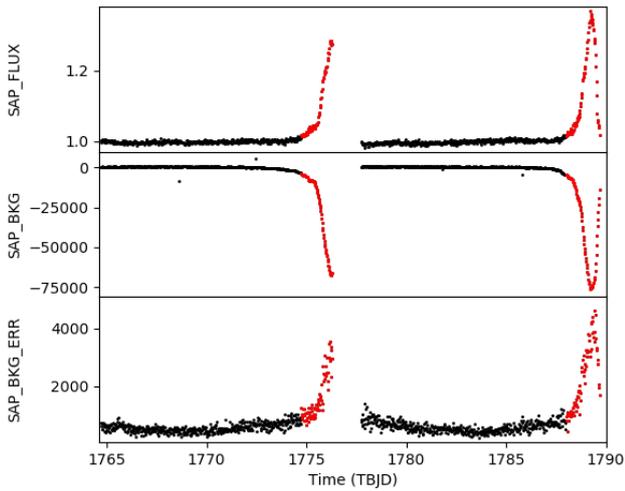

**Figure 9.** Example of the background error cut being applied. Flux data is cut in regions where the background error exceeded two times the median background error, as highlighted in red.

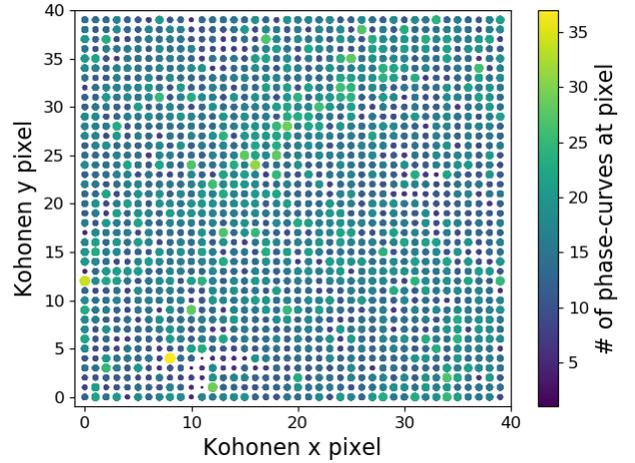

**Figure 10.** S17 QLP SOM after applying background error clean. This displays a much cleaner distribution across the SOM map and traditional EA/EB hot-pixels, compared to the systematics-dominated original S17 SOM in Figure 5.

Hence by clustering similar shapes of phase-curve close together, the SOM effectively gives a visual overview of any common systematics in the data-set which may otherwise be missed. It also gives the user a useful set of trial light-curves to test techniques of removing these troublesome systematics, without having to check each by eye. Furthermore, the clustering gives an overview of which systematics appear to be related in shape, and can more easily allow the user to view what is being removed in any related principal component analysis. Identifying and removing these systematics one by one can significantly aid searches for planetary transits, whether the stars are young or otherwise. The SOM's ability to identify these automatically and visually thus make it attractive as an additional tool for narrowing down troublesome systematics. This analysis also illustrates the necessity of carefully checking individual light-curves when applying wider machine-learning techniques to the *TESS* data, as some phase arrays can be suggestive of real activity when the extra information given in the original light-curves is simplified.

### 4.3 Self-trained vs Pre-trained SOM

The are two primary ways to train the SOM, both of which have different benefits to understanding the variability in the young star population. We here define these two ways as 'Self-trained', where the SOM is trained directly on the young star light-curves analysed in this work, and 'Pre-trained', where light-curves from known variability groups are used to train the SOM. In the latter case the young star light-curves are then mapped onto the trained SOM in order to see where they fall in regards to the known classes of variability. The prior training method is used for the majority of this analysis, as it is useful to explore the inherent variability in the young star light-curves specifically. However, the 'pre-trained' SOM may be useful for future machine-learning classification of these light-curves, so is also explored here.

Self-training the SOM is the easiest option for training the SOM, and is particularly useful as it does not require any prior knowledge about the expected shape of the light-curves, or indeed any extra data-set. In this method the SOM is trained by inputting the young star SOM arrays prepared in Section 3 directly into the SOM algorithm. The algorithm then moves these light-curves into similar regions when they share similar shapes, as explained in Section 4.1. The final Kohonen map in this instance is thus trained on the shapes present only in the young star light-curves, so can be useful to explore the overall variability trends and any clusters of interesting young star variability. Results from this training method are explored through most of Section 5, and in the systematics discussion above.

The alternative approach of pre-training the SOM is perhaps most useful for classification/grouping of light-curves into pre-known classes of variability, but can also be used as a valuable way to compare different data-sets on a common and repeatable structure. This method involves training the SOM on a pre-selected set of light-curves (often pre-classified into variability classes) and then mapping all new light-curves onto the same pre-trained Kohonen layer. This is helpful because it breaks the randomness of each newly trained SOM, and means that the Kohonen shape at each pixel of the SOM will remain unchanged, even when new light-curves are mapped onto it. This makes it very useful for comparing data-sets (e.g. different *TESS* sectors), as each region of the final SOM map will correspond to the same shape, so only the number of light-curves at each spot will vary between runs. Note that Kohonen (1982) discussed an alternative approach to compare data-sets whereby a known 'seed' is used in the SOM initialisation to ensure that the distribution of shapes on the SOM is repeatable between SOM runs, but this was not tried in this work.

In this case light-curves from the pre-classified K2 Variability Catalog (Armstrong et al. 2016) were chosen to pre-train the SOM, as these would allow specific SOM regions to correspond to known variability classes after training. Stars in the K2 Variability Catalog were classified using a combination of a Self-Organising Map and a Random Forest classifier, making them well-suited to use in further SOM analysis. In total the catalog consists of 68910 classified light-curves, including 154 ab-type RR (RRab) Lyraes, 377 $\delta$ Scuti stars (DSCUT), 133 $\gamma$ Doradus variables (GDOR), 183 Algol-type eclipsing binaries (EA), 290 contact eclipsing binaries (EB) and 9399 other periodic variables (OTHERPER) of high-probability (P>0.7) in campaigns 0-4 of the K2 mission. Ideally *TESS* light-curves would be used for each of these stars to maximise similarity to the young star data-set, however at the time of writing only 8903 of these stars had been observed by *TESS*, as the K2 mission took place primarily on the ecliptic plane, where *TESS* observations are still ongoing. Instead the Warwick detrended light-curves from (Armstrong et al. 2016) were chosen as the primary data-source, allowing the primary rotation periods derived in this source to be used directly. SOM arrays were then built in the same manner as outlined in Section 3, using the periods presented in Table 5 of (Armstrong et al. 2016). Two separate approaches were then used to train the SOM: firstly the SOM was trained using all light-curves except those classified as 'Noise' (24,947 light-curves) and secondly using all 68,910 light-curves at once. The resulting SOMs are shown in Figures 22 and 23 respectively and are discussed in detail in Section 5.4.

# 5 RESULTS

## 5.1 Overall results

The overall SOM built and trained on all young star CDIPS light-curves from the first year of *TESS* data is shown in Figure 11, with an overview of the main features presented in Table 1. The raw result from the SOM is somewhat subtle, as the most important information is hidden in the pixels themselves (a sample of which are shown alongside). However, there are still some important features to note in the overall SOM shape. For clarity, the pixels have been colour-coded and given radii in the overall SOM plot based on the number of light-curves (or their respective phase arrays) which ended up at that pixel after training on the CDIPS young star data-set. This aids the eye to see clusters of larger numbers of light-curves in lighter, denser regions. It is important to remember that the SOM boundaries wrap, such that pixel [0,0] is diagonally adjacent to [39,39]. This can clearly be observed in the continued 'ridge' from [0,13] to [39,13] and in the specific phase-curves at the edges of Figure 11.

The most striking features in this SOM are the hot spots at [6,2] and [8,6], represented as the yellow spots on the SOM, alongside the slower evolution of light-curve shape across the remainder of the map. The SOM pixels for these two 'hot-pixels' at [6,2] and [8,6] are plotted in Figures 12 and 13 respectively, and appear to represent two very well-known classes of photometric variability: Beta Lyrae type contact eclipsing binaries (EB; [6,2]) and Algol-type detached eclipsing binaries (EA; [8,6]). Querying specific light-curves at these pixels confirms this classification, with clear examples of both type of binary evident. This suggests that this clustering method is very effective at finding eclipsing binaries, and further gives an efficient and intuitive view of those present in the data-set. A closer examination of the EA-like signals is also useful given that transiting planets could give very similar eclipse shapes, especially for those pixels where secondary eclipses are not obvious in the phase-curves.

The regions immediately around these 'hot-pixels' are interesting to examine, representing shapes similar to the most distinct EA and EB eclipsing binaries. The region surrounding pixel [6,2] is comparatively sparse, suggesting that it is a very distinctive shape, with few variations from the shape present. On the other hand, there are a number of warm pixels to the left of [8,6], which when queried show evidence of a second eclipse, for which a hint can be seen on the bottom of Figure 11, at pixel [5,5]. This comparison suggests that there is considerably more shape variation in EA light-curves viewed in these young stars compared to the variation observed in EB light-curves, much of which may be down to the period of the detached EAs. To the right of both of these hot pixels is a steep drop-off, which makes sense given the distinct change in shape to noisy sinusoids which occurs towards pixel [15,5] (also pictured in the bottom panel of Figure 11).

A first glance at the full SOM distribution through individual pixels reveals the rich array of variability present in young star light-curves. While sinusoidal pixels are certainly common (as is expected given the common spot modulation in young stars), it is not hard to find alternative pixels where simple sinusoidal detrending would be completely inappropriate (e.g. [10,30] and [30,20]). It is also obvious that in some regions of the SOM the shape trend varies slowly (e.g. the diagonal from [0,10] to [20,30]), but in others the changes in shape are rapid (e.g. the swift change between pixels [10,30] and [15,35]). When analysing the wider SOM distribution further, two key things to look out for are light clusters of the most common light-curves (e.g. [2,32] and [15,5]) and dark regions (e.g. [21,26] and adjacent to [6,2]) separating quite distinct shapes of variability. As discussed previously, it is these individual clusters and not the overall SOM distribution that are of greatest importance, especially as the specific locations of clusters will change with every run without pre-training. The most common phase-curve shapes aside from the EA/EBs discussed previously appear to be variations on the asymmetric sinusoids at [2,32] (similar to [5,30] in Figure 11) and negative sines/cosines (e.g. the distinct diagonal cluster from [34,7] to [13,24] in Figure 11). However one particularly unusual one is [22,24], falling in the middle of another less-dense region, like the EA and EB pixels. Querying the individual phase-curves and corresponding light-curves (e.g. Figure 14) at this point curiously reveals most to contain flares, suggesting that this extended peak in the top panel of Figure 14 is caused by averaging the different phases of maximum flare brightness.

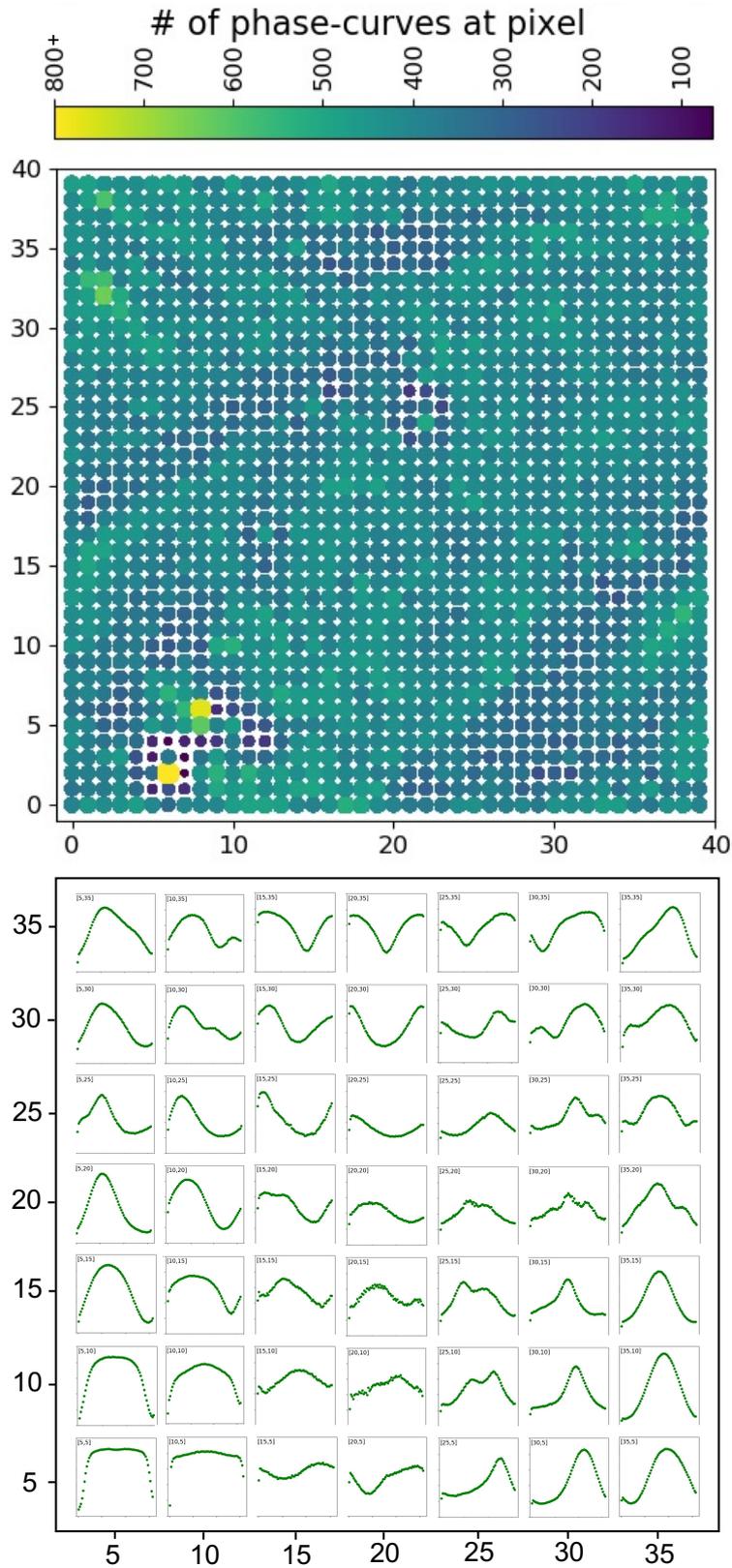

**Figure 11.** S1-13 Final CDIPS SOM. *Top*: Final SOM map for all CDIPS light-curves, showing their clustering and distribution into the 40x40 pixel map. The number of individual light-curve arrays at each point is represented both by colour (see the colour-bar above), and by pixel radius. The most common pixels are thus lighter colour and in regions of the map with less white space. *Bottom*: Plots of individual SOM pixels from the trained SOM, at 10-pixel intervals over the SOM, to show the approximate evolution of light-curve phase arrays over the SOM map. Note that the SOM wraps at all edges, so the shapes at y=39 are largely very similar to those at y=0.

**Table 1.** Overview Table of full S1-13 Self-Organising Map

| Variability Type | Main Position |
|---|---|
| Algol type eclipsing binaries (EAs) | [8,6] |
| Beta Lyrae type eclipsing binaries (EBs) | [6,2] |
| Flares | [22,24] |
| Dipper-like behaviour | Regions around [10,30], [25,16], [35,1] and [35,29] |
| Main sinusoidal modulation | Distinct diagonal band North-East from [34,7] to [13,24] |
| Asymmetric sinusoids | Region around [2,32] |
| Other periodic signals/shapes | Large region around [30,30] |
| Flat/Noise-dominated | Large region around [18,10] |

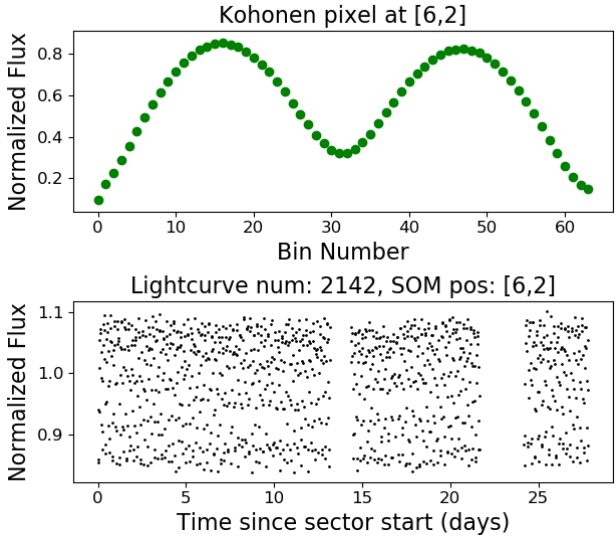

**Figure 12.** Phase curve and example light-curve for Kohonen pixel [6,2] in the final S1-13 SOM, representing the position of most EB signals.

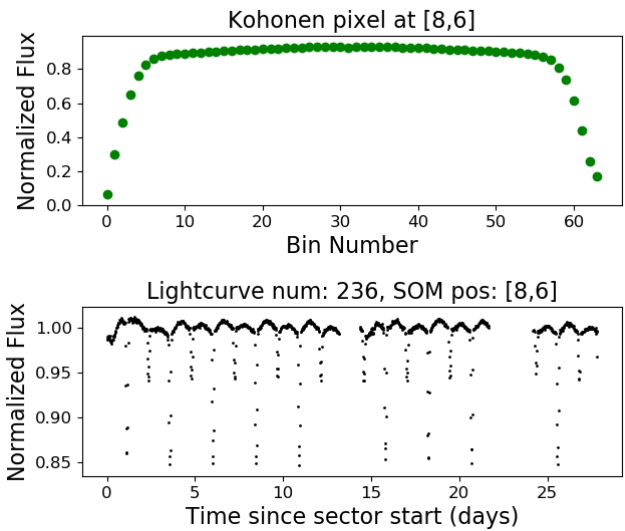

**Figure 13.** Phase curve and example light-curve for Kohonen pixel [8,6] in the final S1-13 SOM, representing the position of most EA signals.

One variability shape that proved harder to find than originally expected was the Dipper/burster class. As these light-curve shapes are thought to be caused by occultations by wraps of accretion columns in circumstellar disks (Hedges et al. 2018), they could theoretically take quite unique and unusual shapes, meaning that they may not all fall on one defined region of the final map. However, through examination of individual light-curves at pixels with unusual Kohonen phase arrays, some dipper and burster-like light-curves were found in the vicinity of [10,30], [25,16], [35,1] and [35,29]. Lists and specific examples of these will be discussed in YOUNGSTER III (Battley et al, in prep) after additional supervised machine-learning methods are combined with the SOM to automatically weed out the large number of noisy light-curves also present at these pixels. The co-location of the noisy and dipper-like light-curves at these pixels is an unfortunate side-effect of swiftly-evolving dippers and some heavily noise-dominated light-curves appearing similar when condensed down to 64 flux bins. Considering non-periodic statistics alongside the SOM shapes in the future should help to separate these signals much more effectively.

The remainder of the final map is mostly split into two main large regions, centred on [30,30] and [18,10] respectively. The first region (around [30, 30]) typically contains phase-/light-curves with other periodic variability (similar in vein to the 'OTHERPER' class defined by Armstrong et al. (2016)), appearing as repeated signals with shapes unlike typical sinusoids. Such profiles may be caused by such processes as stellar oscillations or rotating star spots. Some of these also harbour significant evolution even along the duration of a single *TESS* sector. Meanwhile the region around [18,10] consists primarily of flat or noise-dominated light-curves, with no obvious variability. This would be the bulk of a SOM completed for an older population, yet is comparitively rare for such young active light-curves.

A final interesting feature of some of the phase-curves pictured in Figure 11 is that there is some variation in the range of flux, from 0 to 1 for the strongest signals (e.g. [8,6]), down to as low as 0.15 to 0.3 for some of the noisier light-curves (see [22,24], Figure 14). This is somewhat surprising given that all phase-arrays were normalised between zero and one in their construction. Such a discrepancy is explained through examining individual phase-curves at this pixel, showing that many of the light-curves grouped at these pixels are representative of all those light-curves within it, which was seen to distort the main EB pixel shape compared to the 40x40 run. This size of SOM is thus most useful for getting an idea of the overall larger variability classes present in the model by eye, but is considered slightly too simple for representing the wide variation in light-curves shapes present in young star light-curves.

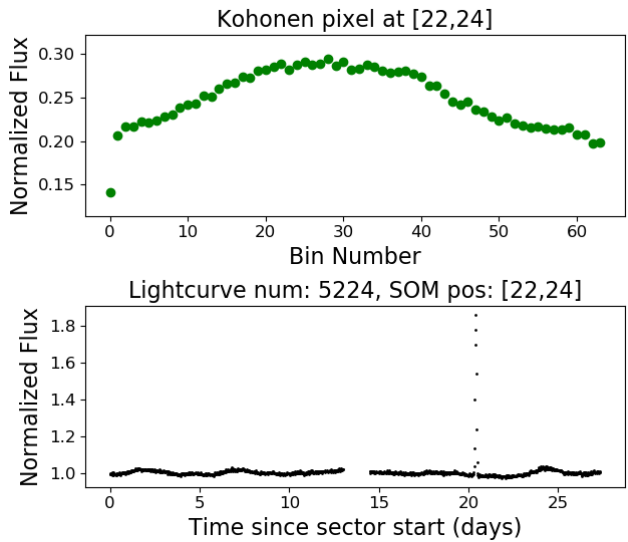

**Figure 14.** Phase curve and example light-curve for Kohonen pixel [22,24] in the final S1-13 SOM, another site of unusually distinct variability.

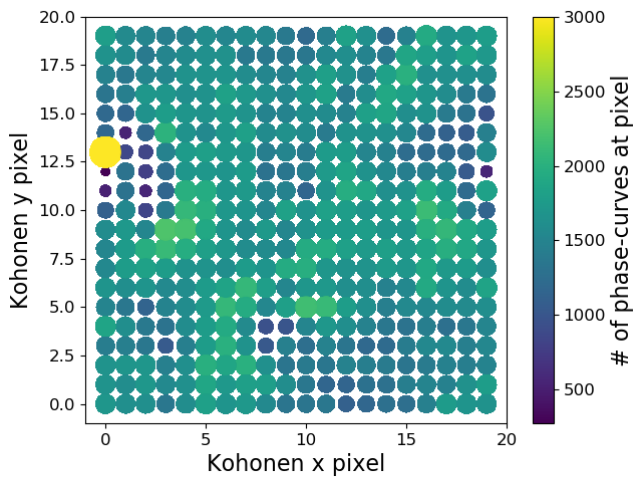

**Figure 15.** S1-13 CDIPS SOM - 20x20

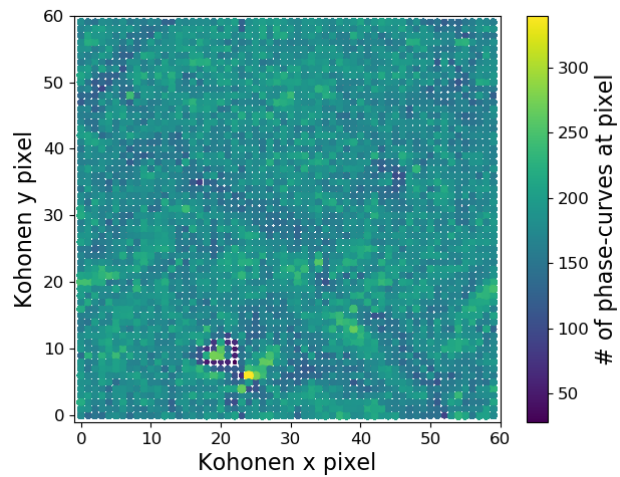

**Figure 16.** S1-13 CDIPS SOM - 60x60

On the other hand, the 60x60 SOM (Figure 16) exacerbates individual variability types while sacrificing the ease of understanding overall trends by human eyeballing. Examining all 3600 pixels by eye is unwieldy, and unlikely to give much of an understanding of the large-scale variations in the data-set. It is the hot-pixels where this larger SOM really excels however, showing considerably more small-scale peaks and troughs than the 40x40 SOM. As in the 40x40 case, the key features to look for are bright pixels surrounded by darker pixels, as these are comprised of arrays very different to the surrounding data. The clearest example of this is around the pixel [20,9] in Figure 16, which once again represents the distinctive group of Beta-Lyrae type Eclipsing Binaries. Slightly subtler interesting groups can be found at pixels such as [37,0] and [40,30], pictured in Figures 17 and 18 respectively. While many phase-curves ending up at these pixels are simply due to noise scatter, others (such as those pictured) are worth investigating further for behaviour such as dippers, bursters, dusty systems and interesting variability.

### 5.3 Sector by sector analysis

As well as generating an overall SOM for the complete Year 1 CDIPS light-curves, SOMs were generated for each *TESS* Sector individually. Because the constructed phase arrays are of a constant size and normalised between zero and one, direct comparisons between sectors are easy, despite variations in data gaps and periods of extra scatter. Note however that as most young stars are concentrated along the galactic equator, the number of targets per *TESS* sector was found to vary considerably, ranging from as few as 3120 in Sector 3 to 146557 in Sector 11. Because of this, building a SOM out of Sectors 1-5 individually was not found to be very instructive, so instead these were combined into a single S1-5 run totalling 26,153 targets.

Alike to the full Year-1 SOM run, the self-training method of building the SOM was used for each sector, with the results for sectors 1-5 shown in Figure 2 and the remaining sectors presented less similar in shape to each other, hence average out to smaller ranges.

### 5.2 Variation in SOM size

Two additional sizes of SOM were trialled to yield more information from the SOM: 20x20 and 60x60, both of which highlight different aspects of the young star data.

The 20x20 pixel SOM (Figure 15) provides an easier way to see overall trends in the data-set, including the main evolution from sinusoids (e.g. [3,9]) to more complex shapes (e.g. [10,0]) and the 'hot-pixels' at [0,13] and [19,11] for the EAs and EBs respectively. This makes it a useful first overview to the data-set, and can capture any significant systematics present. However, given the much smaller mapping space to spread things out, it is prone to combining slightly different types of variability into a single pixel. Such a combination typically makes the individual SOM pixels less in Figures A1 to A8. Although specific positions vary because of the random initialisation of the SOM, the major characteristics of the final maps for each sector are very similar, with hot-pixels for the EA/EB eclipsing binaries surrounded by the lowest density regions, alongside more steady evolution over the rest of the map. Some discrepancies do however exist with regards to additional hot pixels, such as pixels [6,22] in the S1-5 SOM and [7,14] in Sector 9. In the case of pixel [6,22] of the Sector 1-5 SOM, this pixel represents a noisy sinusoid (plotted in Figure 19). This is thought to be caused by the extra pointing noise present in the early *TESS* sectors, which were corrected in Sector 6 onward. Because these light-curves represent a reasonably small number in the overall Year 1 data, these light-curves do not form a large enough group to be a significant hot pixel in the full Year 1 run.

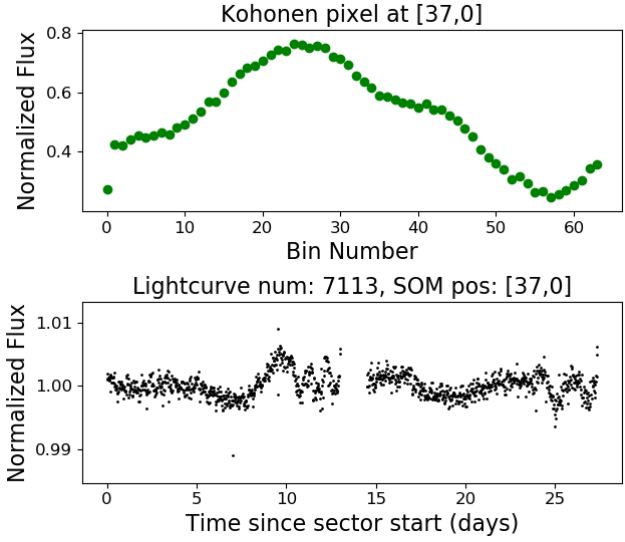

**Figure 17.** Example 1 of interesting Kohonen-pixel shape and corresponding example light-curve from 60x60 SOM run - pixel [37,0]

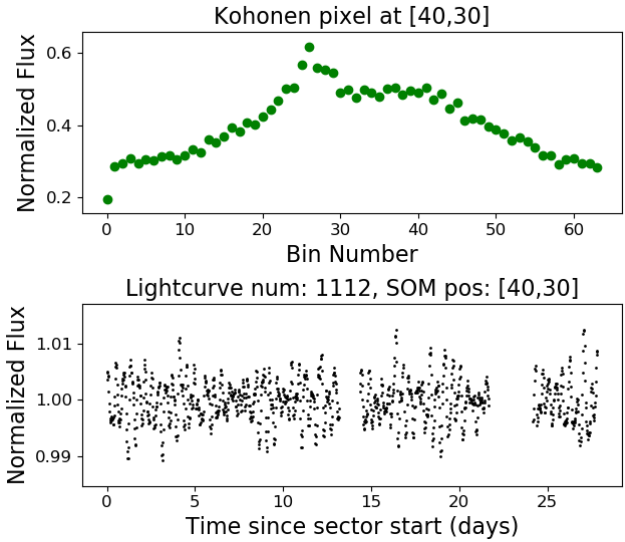

**Figure 18.** Example 2 of interesting Kohonen-pixel shape and corresponding example light-curve in 60x60 SOM run - pixel [40,30]

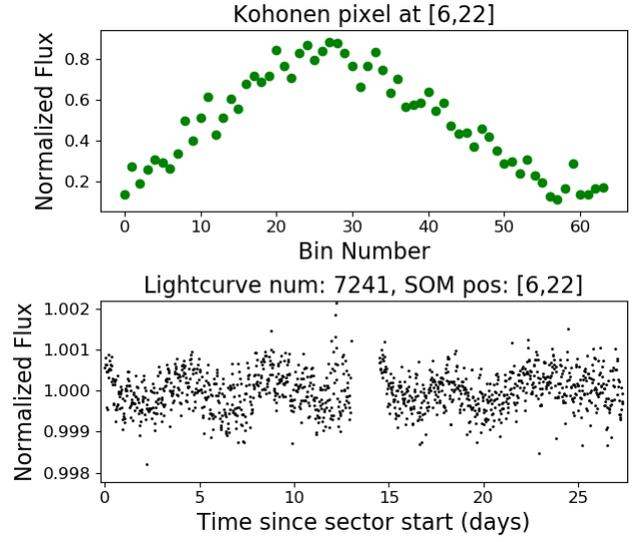

**Figure 19.** Phase-curve and example light-curve for additional 'Hot-Pixel' [6,22] in S1-5 SOM, showing noisy sinusoidal geometry.

Meanwhile, the additional two hot pixels in the Sector 9 plot at [7,14] and [35,10] have the geometry plotted in Figures 20 and 21 respectively. These are less easy to characterise using the SOM pixels alone, and hence are better understood by checking some of the individual light-curves at these points. Closer inspection of these light-curves revealed that the targets at both pixels were dominated by excess noise, coupled with a distinctive background signal consisting of V-shaped dips at the beginning of both orbits. An additional flux ramp before these V-shaped dips in the light-curves at [7,14] changed the overall shape at this pixel further. All of the light-curves grouped to these pixels came from suspiciously similar light-curve numbers (Sector 9 light-curves 45,000-50,000 for pixel [7,14] and around 1400-4000 for pixel [35,10]), suggesting that they come from similar regions of the CCDs originally. Sector 9 also contains some very crowded regions of sky near the ecliptic plane, which may explain why so many light-curves were adversely affected. This example illustrates the care needed when analysing the final product of the SOM, showing that it is always best to double-check the original light-curves for any particularly interesting Kohonen pixels.

Nonetheless, the similarity in SOM clustering and overall trends present between sectors show that the SOM remains effective across all sectors, with the one proviso that enough light-curves are available to allow significant clustering to occur. This bodes well for using SOM positions as a useful metric for further machine learning classification of young stars, as useful results can be obtained even when only running a single month of data.

## 5.4    Pre-trained SOM

Results from pre-training the SOM on the K2 Variable Catalogue data (Armstrong et al. 2016) are shown in Figures 22 and 23, excluding and including those light-curves classified as 'noise' respectively. The same trained SOM underlies both the left and right hand side of each figure, with the K2 Variability Catalogue light-curves mapped onto the SOM on the left and the full *TESS* S1-13 young star light-curves mapped onto the SOM on the right. This means that the shape of each pixel in the K2 Variability Catalogue map on the left corresponds to the same shape at the corresponding pixel on the right, allowing direct comparisons between expected variability classes and the position of young star phase-curves.

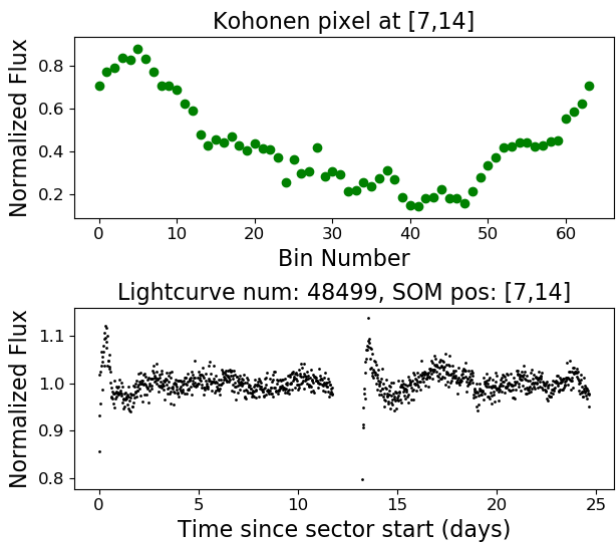

**Figure 20.** Phase-curve and example light-curve for additional 'Hot-Pixel' [7,14] in S9 SOM. Inspection of individual light-curves revealed this to be caused by the distinctive background signal than can be seen at the start of each orbit (steep up then u-shaped dip) in the light-curve.

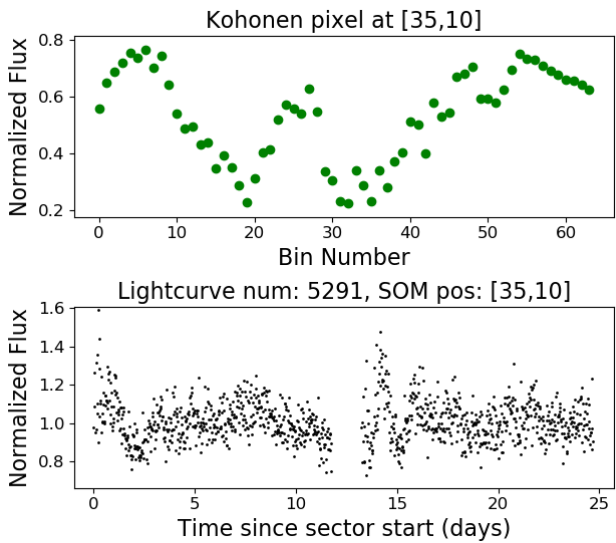

**Figure 21.** Phase-curve and example light-curve for additional 'Hot-Pixel' [35,10] in S9 SOM. Similar to [7,14] this was caused by additional noise and a distinctive background signal.

In the case of the 'no-noise' SOM (Figure 22), the plotted variability classes were observed to cover most of the 40 x 40 grid, with a wide diagonal band of primarily other periodic (OTHERPER), delta-scuti (DSCUT) and Gamma-Dor (GDOR) variables. Meanwhile eclipsing binaries of both types (EA/EB) congregated around [39, 30] (remembering that the SOM wraps at the each edge), distinctly off-set from the other types of variability. In this case the white space left on the plot is mostly full of targets with class probabilities of <90%, representing those with less clean-cut classifications from the K2 Variability Catalogue. Interestingly, when the *TESS* S1-13 CDIPS light-curves analysed in this work were applied to this same map, the vast majority of them fell on regions of the map with low densities of K2 variability catalogue targets. This suggests most of the shapes seen in the young *TESS* light-curves (likely from alternative types of variability or noise) are different to the specific variability classes defined in the K2 variability catalogue. Such poor classification performance stresses the importance of conducting young-star specific classification rather than simply relying on previously known variability classes. It should be noted however that some of the young star light-curves do still fall on, or nearby, known variability classes on the 'no-noise' SOM, particularly over the centre of the main OTHERPER block of targets and adjacent to the main group of eclipsing binary light-curves. Querying CDIPS young star light-curves which appeared in this EB-adjacent region does reveal them to be eclipsing-binary-like signals in the *TESS* data as well, suggesting that these types of signals are the ones which are best handled by (and most similar to) the K2 Variability Catalogue trained SOM.

Examining the SOM trained on the full K2 Variability Catalogue (Figure 23) provides more promising results. Because training of the SOM spreads variability across the entire map, there is considerably more white space in this latter case, where most of the 'Noise' light-curves fell. However, the OTHERPER sources are still spread out over a large proportion of the SOM, suggesting that there is a lot of variation in their specific light-curves, compared to the much more localised EA and EB light-curves. In this particular run the eclipsing binary signals were split into two main groups rather than a single region, but while there is some difference in the shape of the phase-curves between these groups (with the top right group having wider dips), this is considered more to do with the random initialisation of the SOM than significant differences in shape, as a previous run with the same K2 Variable Catalog data placed all EAs/EBs in a single joined group. The inclusion of the 'noise' class appears to aid the SOM performance considerably, both by grouping light-curves similar to known variability classes more obviously and by allowing internal structure and specific groups of the light-curves which fall onto quieter regions of the plot to be viewed. Of particular interest is the main EA/EB group at [19,25] on the K2 Variability Catalogue map, which is clearly reproduced as an area of high density in the young star map on the right. As is expected, querying individual phase- and light-curves ending up at this point reveals true EA/EB-like signals, showing that the SOM is working effectively at finding these. A similar region of high density can be seen at [26,33] in the young star map on the right, just slightly below and to the left of the expected position ([28,35]) in the K2 Variability Catalogue map.

Meanwhile there are a handful of clear over-densities in the quieter regions of the original K2 Variability Catalogue map, specifically around [5,20], [10,28], [16,10], [21,36] and [35,1]. While the shape of the average phase curve at each of these pixels varies somewhat, all of these phase-curves have y-ranges of about 0.3-0.7 rather than 0-1 as they were initialised with, suggesting that they contain more variation between each phase curve at these pixels. This is common where initial light-curves are noisy, causing large variation between each corresponding phase-curve. Indeed querying individual *TESS* phase- and light-curves which congregated at these pixels (e.g. see Figure 24) reveals them to be noise-dominated, significantly more noisy than is usually seen in light-curves extracted from the *K2* mission. This was not entirely unexpected given *TESS*'s reduced photometric precision compared to the *Kepler* spacecraft, as indeed

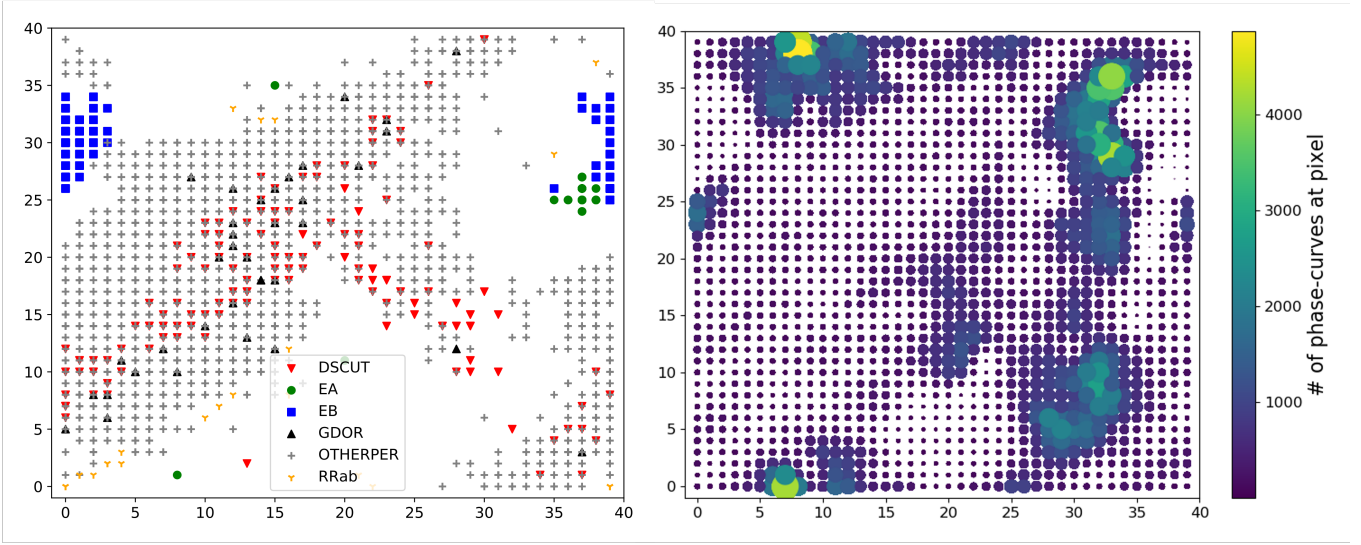

**Figure 22.** SOM trained on the 24,442 light-curves from the K2 Variability Catalogue which had classifications indicating that they were not noise. As usual, axis numbers correspond to Kohonen x and y pixel numbers, similar to Figure 1. Note that only targets with >90% class probability in the catalogue are plotted for clarity. *Left:* The 24,442 non-'noise' light-curves mapped onto the SOM, showing distribution of known variability types. *Right:* All young star CDIPS light-curves from the first year of the *TESS* mission (Sectors 1-13) mapped onto the same K2 Variability Catalogue-trained SOM. The bulk of the CDIPS light-curves can be observed to fall into gaps in the K2 Variability Catalogue map.

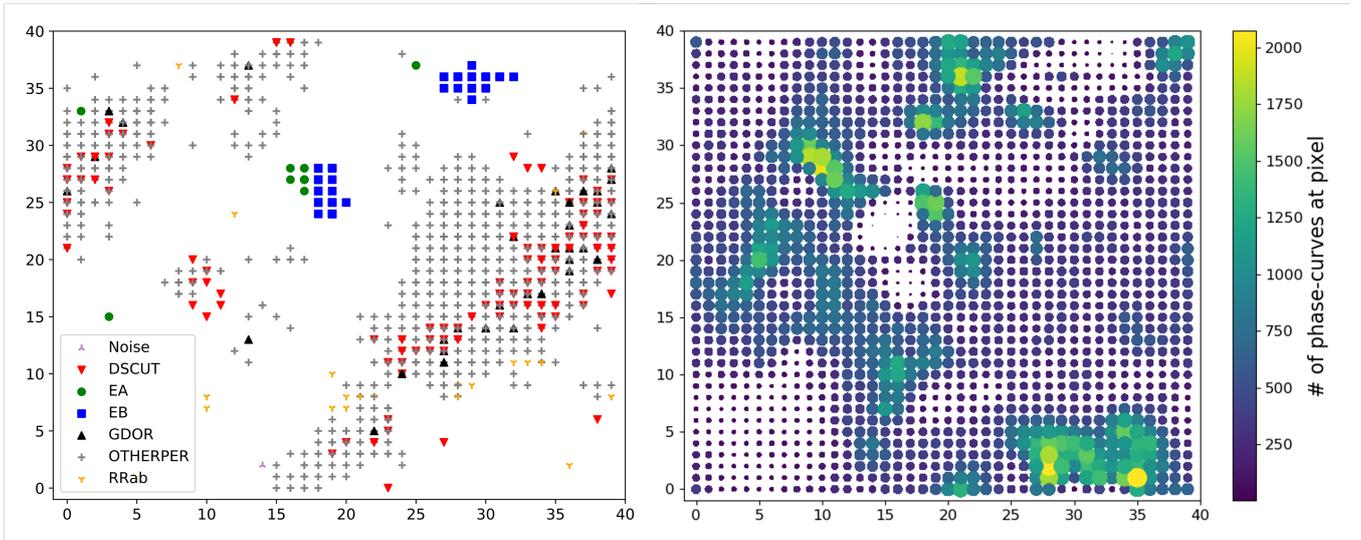

**Figure 23.** Same as Figure 22, but considering all 68,910 in the K2 variability Catalogue, including those classified as 'noise'. Axis numbers represent Kohonen x and y pixels, alike to Figure 1. While overall classification performance is improved (see for example the clear EA/EB over-densities at [19,25] and [26,33]), many light-curves still fall in gaps in the K2 Variability Catalogue trained SOM, likely due to excess noise in the *TESS* light-curves and additional types of young variability not present in the K2 Variability Catalog

.

similarly large noise differences were also seen by Battley et al. (2021) when comparing a number of *Kepler* planets reobserved by *TESS*. Such a problem is particularly challenging for dimmer *TESS* targets, where the instrumental white noise far outweighs any potential variability signals. In order to explore this aspect further, a brighter subset of the CDIPS targets were selected by choosing those stars brighter than Tmag = 12 in the TESS Input Catalog (TIC v8; Stassun et al. 2019). Applying the SOM trained on the full K2 Variability Catalog (including the 'Noise' class) to the phase-curves for these targets resulted in the final map pictured in Figure 25. Through comparison to Figure 23, it is clear that these brighter TESS targets align much better with the previously known variability groups from the K2 Variability Catalog (plotted on the left of Figure 23) than the full CDIPS light-curves did, with the bulk of the targets falling on the main band of OTHERPER/DSCUT variables or EA/EB clusters. Furthermore, the distinct over-densities observed around [5,20], [10,28], [16,10], [21,36] and [35,1] have all reduced significantly in relative strength compared to the main clusters of known variability type.

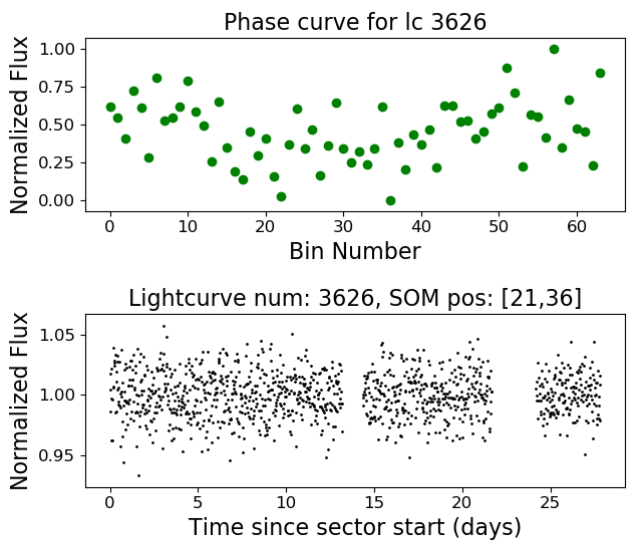

**Figure 24.** Example of one of the noise-dominated *TESS* light-curves (top) and phase-curves (bottom) placed into gaps in the SOM pre-trained on the full K2 Variability Catalogue.

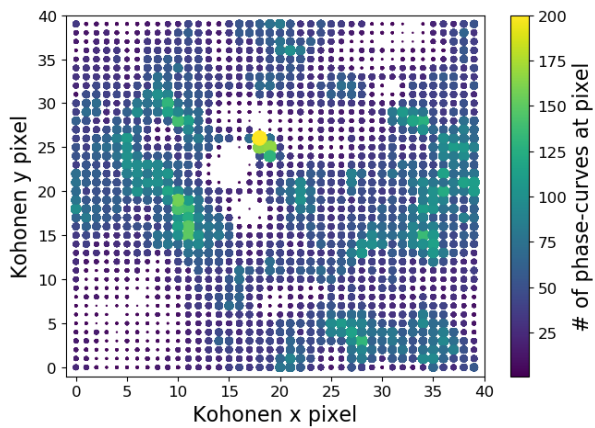

**Figure 25.** Bright (Tmag<12) CDIPS targets from Sectors 1-13 mapped onto the full K2 Variability Catalog trained SOM shown in Figure 23. Although a few smaller areas of higher density still exist in gaps in the original K2 Variability Catalogue SOM (e.g. on the lower right of the plot, around pixel [30,3]), the majority of the data now falls into regions of previously known variability classes.

None the less, not all of the light-curves falling in the gaps can be explained by the excess noise in *TESS*. Alongside these phase-curves exist a number of unusual variability types, including dipper-like behaviour and unusual rotation-like signals. It is these phase curves that remain in regions like [28,3] and [22,36] in the brighter CDIPS SOM (Figure 25). The reason for these falling in gaps could be partially explained by the included young stars having a much wider sample of young ages compared to the often older K2 stars. This would mean that some K2 light-curves may have types of variability not seen in young stars, and some particularly young stars may have types of variability not present in bulk in the K2 Variability Catalogue. Additional differences in shapes could also be explained by the differences in *TESS* sector lengths (roughly 27d), compared to K2 campaign length ( 80d for the analysed 0-4th campaigns), meaning that K2 targets may evolve more in phase over the time period. However, such interesting types of variability will be much more easily explored and explained when not overwhelmed by the large number of noisier *TESS* light-curves also constrained to these pixels.

These challenges show that although pre-training the SOM shows some promise in classifying new targets into previously known variability groups, the performance of this classification would benefit significantly from using light-curves from the same survey/instrument compared to archival data. In future analyses in the YOUNGSTER program, *TESS* light-curves will be used instead, based on objects with previously classified variability classes. Unfortunately at the time of writing not enough of these existed to test such an analysis, however with *TESS* now observing the ecliptic plane, many of these pre-classified targets from the K2 Variability Catalog should soon receive *TESS* data.

### 5.5    Catalog of Eclipsing binaries and eclipse events in TESS Year 1

As discussed in Section 5.1, the SOM method appears to be very effective at highlighting and separating the distinctive shapes of eclipsing binaries from other variability present in the young star light-curves, appearing as distinctive hot-pixels in all the self-trained SOMs generated in this work. As young eclipsing binaries are very important to understanding the early evolution of star systems, those identified in this work are presented here in the hope that they will be useful for the wider young star community. Although the bulk of these light-curves ended up at the hot-pixels at [6,2] and [8,6] in the complete Year 1 SOM (Figure 11), there are additional binary-like signals in pixels adjacent to these, especially to the left of the main detached eclipsing binary (EA) pixel at [8,6]. Because of this, a more conservative method of selecting light-curves was used, selecting phase-curves by using a cut based on the Euclidean distance from the archetypal phase-curve at the strongest pixels. The specific cuts used were $d_{xy} < 1.56$ from pixel [8,6] for the EA candidates and $d_{xy} < 1.37$ from pixel [6,2] for the EB candidates. This allowed the most similar light-curves from the surrounding area to be selected whilst reducing the number of false-positives that would be selected by simply choosing all pixels which harbour some eclipsing binary-like signals. These cuts resulted in a total of 13,366 candidate EA signals and 2448 candidate eclipsing binary signals, including many systems which were picked up in multiple sectors. All selected light-curves were then checked by eye in order to catch any remaining obvious false positives. Such false positives typically consisted of excess noise, sharp stellar activity or obvious blends with other nearby targets. Following this step and a uniqueness cut on TIC ID, 8103 EA and 1297 EB candidate systems remained, all of which are presented in Table 2. This significant drop is largely not due to mis-classifications by the algorithm, but instead light-curves with the EA/EB shape which were subsequently found to be blends during human eyeballing, or those which appear in multiple sectors. As such, some care is advised when interacting with this catalog, as further less obvious blends may still exist due to the large *TESS* pixels. This catalog is not exhaustive, but represents the cleanest sample of young eclipsing binaries identified by the SOM analysis of these young stars. The catalog will be updated with less distinctive cases using future supervised machine learning methods in the continuing YOUNGSTER program.

**Table 2.** Candidate eclipsing binary signals found in Year 1 of the CDIPS *TESS* light-curves. Light-curves with only single eclipses are labelled as 'Single' in the Period column. A machine-readable version of the full table (9400 targets) is included alongside the paper online. Periods shown are the primary period found in section 3.

| Gaia DR2 ID | TIC ID | Period [Days] | Type |
|---|---|---|---|
| 6885253422567909760 | 197601231 | Single | EA |
| 6619768913628810624 | 32197062 | 0.331 | EA |
| ... | ... | ... | ... |
| 6630166342263764992 | 365770318 | 0.289 | EB |
| 5777534485542148608 | 310104320 | 1.17 | EB |

# 6 DISCUSSION

With these promising results, it is clear how powerful the Self-Organising Map technique is for exploring the variability of young stars, both in terms of viewing the broad distribution in variability and for isolating individual groups of distinct variability such as EAs and EBs. The final Kohonen map provides an intuitive and useful overview of the entire first year of young star data in *TESS* (especially when individual pixels are queried) and can even speed up identification and characterisation of systematic errors in the data. This is the first time that such a technique has been applied to a dedicated sample of young stars, as well as the first time such a technique has been applied in bulk to *TESS* data. Applying this technique to young stars specifically allows for a deeper view of unusual variability types which would easily be drowned out by some of the distinctive variability classes presented in older samples of stars (e.g. Cepheids, RR Lyraes). This is especially important as their spot modulation and comparatively short-period variability can result in significantly different shapes to the standard accepted variability classes. Indeed one particularly interesting feature of the full S1-13 SOM (Figure 11) is how uncommon basic sinuisoidal shapes are, suggesting that simple spot modulation with single large spots dominating the photosphere are not the dominant variability type. This wide array of different shapes helps to explain why transit surveys struggle to find planets around young stars with traditional 'one-size-fits-most' detrending methods.

However, despite the useful results offered by the SOM analysis, some key provisos remain. Most importantly, as the SOM technique works by folding light-curves and binning phase-curves by the strongest periodic signal, it effectively obscures or reduces any short-term or aperiodic activity, including important features such as flares and aperiodic dipping activity. Similarly, because all phase-curves are normalised, some noisy light-curves can have their slight variability greatly exaggerated. These two factors show the need for extra factors like period, peak-to-peak scatter statistics and amplitude measurements when attempting wider classification of light-curves. This is corroborated by Brett et al. (2004) who point out that because real-world examples will have a lot of similar shapes at the edges of each class, additional diagnostic attributes are required for safer, more accurate classification. In order to combat this in the continuing YOUNGSTER program, information from the generated SOMs (in particular the final SOM position and Euclidean distance from final pixel) will be combined with known stellar in-formation, light-curve statistics and known variability classes in a wider random-forest-based machine-learning method in future (see YOUNGSTER III, Battley et al., in prep). While biases and varying membership likelihoods in the CDIPS target-list mean that reliable statistics are not easily obtained from this current sample, using these SOM-based techniques with more accurate target-lists (such as in individual moving groups/associations) have the potential to provide statistical overviews of the shapes present in a specific sample.

Because of *TESS*'s observing strategy, most targets analysed in this study only had a single sector of *TESS* photometry, totalling approximately 27days of observations. This means that young stars in this sample with dominant activity periods of longer than this window will not be folded by their correct period, instead likely appearing as aperiodic. One way to solve this challenge would be to focus on stars in the *TESS* continuous viewing zone (CVZ), however because the majority of young stars fall near the galactic plane, while on the order of 500 stars in the CDIPS sample analysed here fell in this area of the sky, far fewer than necessary to generate a reliable and informative self-organising map. Using new data for these targets from *TESS*'s extended mission will help to extend the observational baseline for these objects, however at the time of writing no light-curves beyond Year 1 of the *TESS* mission were available from the utilised CDIPS pipeline. None the less, it is clear that even using a single sector of data allows a wide distribution of variability shapes to be observed. Indeed, one of the key benefits of using binned phase-curves instead of light-curves as the primary SOM data-input is that light-curves from different sectors and data-lengths can be easily compared. Theoretically this should allow data from different instruments to be compared also, however this was shown to be more challenging than originally expected when comparing to the K2 Variable catalog (Armstrong et al. 2016) in Section 5.4 above. In the future this pre-training will be improved by using *TESS* light-curves for objects pre-classified into known variability classes, alongside potential new groups found through deeper SOM analysis.

One additional aspect of the SOM that was made clear by the sector by sector analysis was the required number of light-curves before notable patterns and groups could be seen on the map. In this case, primarily using a 40x40 SOM, 10,000 light-curves were found to be the point at which clustering became significant, which is why the first five sectors were eventually run as one (Sectors 1-5 had 4772, 3535, 3059, 5435 and 9353 young star light-curves respectively). Naturally smaller SOMs could be used to aid clustering in the case of fewer light-curves, but the 40x40 SOM was found to be a good balance of allowing understanding of the global pattern while not merging too many interesting types of individual variability. Thus here it is recommended that at least 10,000 light-curves are utilised for optimum SOM performance.

A challenge of completing this analysis with *TESS* light-curves compared to previous *Kepler*/*K2* analyses is the increased blending from neighbouring stars due to *TESS*'s larger pixels. This challenge fundamentally means that some objects nearby to targets with strong variability may be miss-classified due picking up some of this adjacent signal. This effect is made even worse for young stars compared to the general stellar population due to many young stars existing in relatively dense birth environments. This challenge is well illustrated by the number of supposed EA/EB light-curves which had to be removed in the eyeballing stage of section 5.5, so classifications from *TESS* data alone should be taken with a grain of salt. However, since one of the primary goals of YOUNGSTER program is to detrend the variability of these young stars to search for new young exoplanets, knowing the dominant shape of the variability of each target is still very important, regardless if this is the true variability class of that target.

In Summary, Self-Organising Maps are a powerful technique to explore and characterise young star variability in large surveys such as *TESS*, allowing both an intuitive understanding of the overall variability evolution across the entire sample and highlighting groups of distinct variability. For this reason it will be folded into the wider supervised machine learning methods of the continuing YOUNG-STER program, combined with knowledge of the light-curve statistics and stellar parameters. The final variability classifications from this program can then be used to inform future detrending during searches for new young exoplanets, for example by building variability models or using the knowledge of the light-curve shapes to tune smoothing parameters. Although this analysis primarily focused on the CDIPS light-curves, the methods discussed in this paper are well-suited to other, often more extensive, light-curve sources such as those generated by TESS-SPOC pipeline (Caldwell et al. 2020b), the Quick Loop Pipeline (QLP Huang et al. 2020) and eleanor pipeline (Feinstein et al. 2019), provided that typical systematics are well characterised for these sources. Furthermore, because the SOM analysis is based on binned phase-curves, it can easily be extended to the main 2min *TESS* data and planned 10min and 300s FFI observations.

## 7 SUMMARY

Here Kohonen Self-Organising Maps (Kohonen 1982, 2001) have been used for the first time on a dedicated selection of young stars from the *TESS* survey, in order to explore the wide-scale variability of these interesting objects. This unsupervised machine learning technique allows light-curves to be grouped according to their dominant shapes, without needing prior knowledge of expected groups and clusters. This is particularly important for young stars, where the community is still discovering new types of stellar activity. All currently available light-curves from the Cluster Difference Imaging Photometric Survey (CDIPS, Bouma et al. 2019) are analysed here, representing over 600,000 young star light-curves from the entire first year of *TESS*'s observations.

Overall Self-Organising Maps reveal young star activity shape distribution to be more varied than expected, with a significant spread over the whole parameter space of shapes, from archetypal sinusoidal signals to more varied 'dipper' or pulsation-like activity. The SOMs are also very proficient at separating eclipsing binary and potential transiting signals from the data, typically appearing as significant hot-spots in the final Kohonen maps. Given the importance of these objects, a target list of those identified in this work is supplied in this work and online. SOMs generated for individual sectors retained similar distributions to the overall Year 1 version, provided that at least 20,000 light-curves were present in that sector (i.e. for all but sectors 1-5, which were combined into a single SOM). This illustrates the SOM's effectiveness in characterising the variability of young stars in a single sector alone, without having to run a full year of light-curve data. While most analysis was completed using a 40 x 40 SOM, 20x20 and 60x60 versions were also tested for the full sample of light-curves, revealing the strongest groups of variability in the first case and a deeper sub-structure of variability in the latter. However, the 20x20 SOM was found to merge too many dissimilar light-curves, while the 60x60 SOM is somewhat unwieldy for in-depth analyses, so the 40x40 version is recommended for future analyses.

Pre-training the SOM was attempted using light-curves with pre-defined variability classes from the K2 Variability Catalog of Armstrong et al. (2016), which proved effective for bright targets but was found to be challenging in dimmer light-curves due to differences between the noise characteristics of light-curves from *TESS* and the *K2* mission. Nonetheless, such a method shows sufficient promise to be worth repeating when sufficient *TESS* data is available for these objects from the *TESS* extended mission.

The self-organising map method was also revealed to give valuable and intuitive insight into leftover sysetematics in the *TESS* data (regardless of which pipeline is used), especially where there is variation in the severity of the effects. This will be particularly useful for diagnosing and understanding the significant scattered light systematic present in *TESS* light-curves for dimmer stars. The increased understanding garnered about young star variability from this analysis will be used to guide future light-curve classification and targeted detrending methods in order to search for new young exoplanets in the continuing YOUNGSTER program.

## ACKNOWLEDGEMENTS

The authors would like to thank the anonymous referee for their comments which improved the quality and robustness of this paper.

This paper includes data collected by the *TESS* mission. We acknowledge the use of public TOI Release data from pipelines at the *TESS* Science Office and at the *TESS* Science Processing Operations Center. Funding for the *TESS* mission is provided by NASA's Science Mission directorate. This research has made use of the Exoplanet Follow-up Observation Program website, which is operated by the California Institute of Technology, under contract with the National Aeronautics and Space Administration under the Exoplanet Exploration Program. *TESS* data were obtained from the Mikulski Archive for Space Telescopes (MAST). STScI is operated by the Association of Universities for Research in Astronomy, Inc., under NASA contract NAS5-26555. Support for MAST is provided by the NASA Office of Space Science via grant NNX13AC07G and by other grants and contracts.

MPB acknowledges support from the University of Warwick via the Chancellor's International Scholarship. DJA acknowledges support from the STFC via an Ernest Rutherford Fellowship (ST/R00384X/1). DLP acknowledges support from STFC and also the Royal Society.

## DATA AVAILABILITY

The full young eclipsing binary list constructed in this work is presented as supplementary material in the online version of this manuscript. All other data generated in this work will be shared on reasonable request to the corresponding author. Light-curves analysed in this work were drawn from public CDIPS and QLP *TESS* data hosted on the Mikulski Archive for Space Telescopes (MAST) Portal (https://archive.stsci.edu/hlsp/cdips and https://archive.stsci.edu/hlsp/qlp).

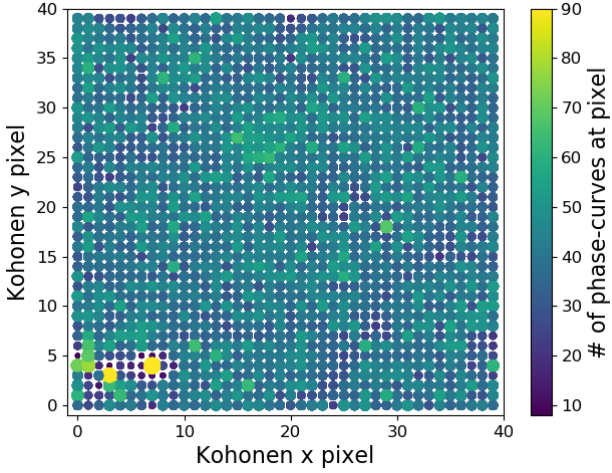

**Figure A1.** Sector 6 SOM

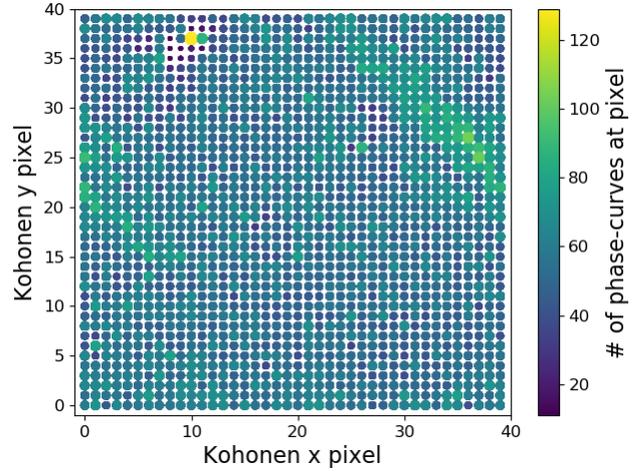

**Figure A2.** Sector 7 SOM

## APPENDIX A:  INDIVIDUAL SECTOR PLOTS

Self-Organising Maps for sectors 6-13 follow for reference. Notes that the SOM created for Sectors 1-5 (which had too few light-curves to be analysed independently) is shown in Figure 2.

This paper has been typeset from a TEX/LATEX file prepared by the author.

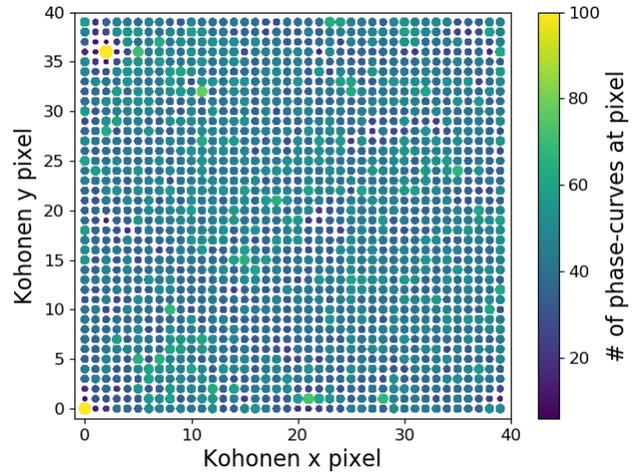

**Figure A3.** Sector 8 SOM

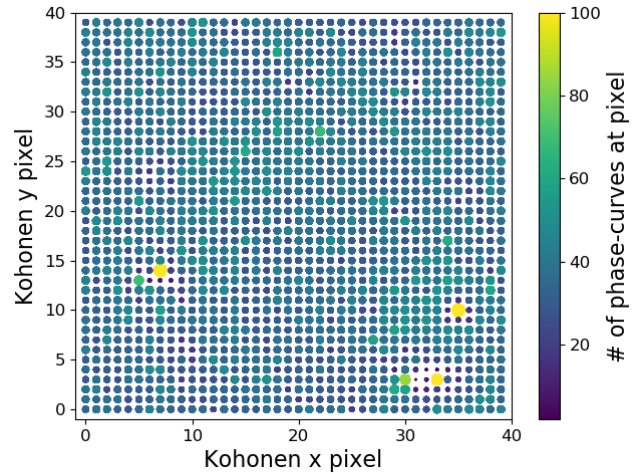

**Figure A4.** Sector 9 SOM

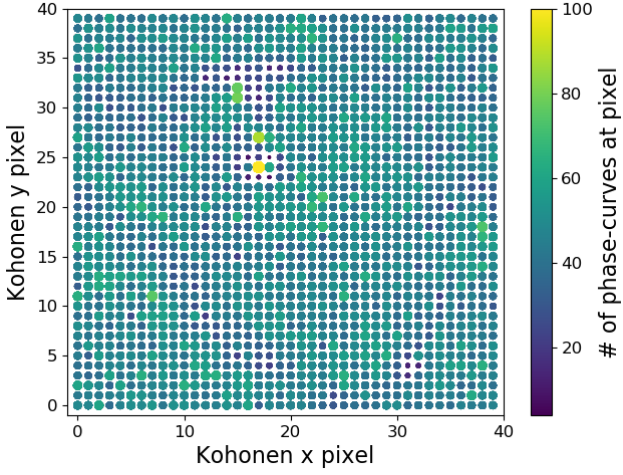

**Figure A5.** Sector 10 SOM

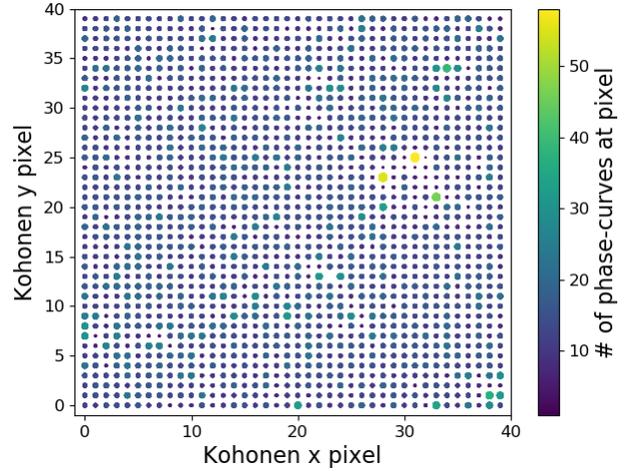

**Figure A8.** Sector 13 SOM

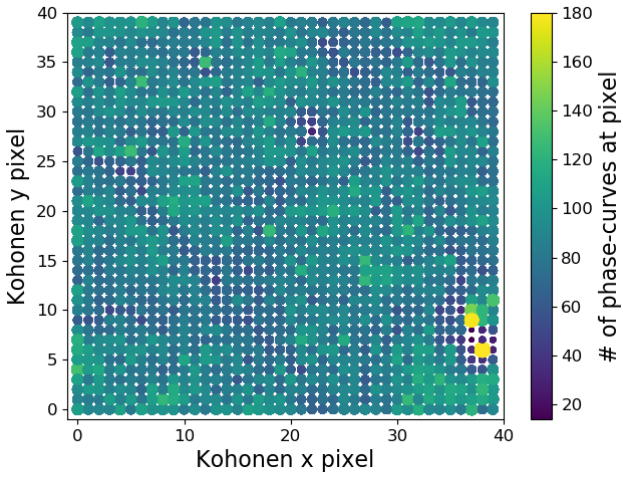

**Figure A6.** Sector 11 SOM

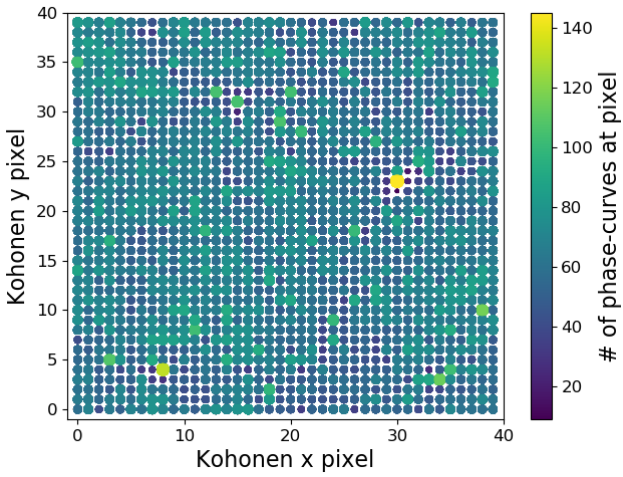

**Figure A7.** Sector 12 SOM